%% file: mnras_template.tex
\documentclass[fleqn,usenatbib]{mnras}

\usepackage{newtxtext,newtxmath}
\usepackage{booktabs,threeparttable}

\usepackage[T1]{fontenc}

\DeclareRobustCommand{\VAN}[3]{#2}
\let\VANthebibliography\thebibliography
\def\thebibliography{\DeclareRobustCommand{\VAN}[3]{##3}\VANthebibliography}

\usepackage{graphicx}	
\usepackage{amsmath}	
\usepackage{caption}
\usepackage{subcaption}
\usepackage[dvipsnames]{xcolor}
\usepackage{multirow}
\usepackage{dirtytalk}

\defcitealias{Dovekie}{Dovekie}
\defcitealias{DES5YR}{DES-SN5YR}
\defcitealias{fragilistic}{Fragilistic}
\defcitealias{Jones17}{J17}
\defcitealias{Hlozek12}{H12}
\defcitealias{Shivvers_2017}{S17}
\defcitealias{2019MNRAS.485.1171K}{K19}
\defcitealias{Vincenzi19}{V19}
\defcitealias{Vincenzi20}{V21}
\defcitealias{BS20}{BS21}
\defcitealias{Wiseman22}{W22}
\defcitealias{DES-CC}{DES-nonIa}
\include{SYSUNCER_VALS}
\usepackage[switch]{lineno}

\title[The Dark Energy Survey: Supernova Reanalysis]{The Dark Energy Survey Supernova Program: A Reanalysis Of Cosmology Results And Evidence For Evolving Dark Energy With An Updated Type Ia Supernova Calibration}

    \author[B. Popovic et al.]{B. Popovic$^{1,2}$\thanks{Email: B.A.Popovic@soton.ac.uk}, P. Shah$^3$, W. D. Kenworthy$^4$, R. Kessler$^{5,6}$, T.~M.~Davis$^{7}$, A. Goobar$^4$, D. Scolnic$^8$, 
    \newauthor
    M. Vincenzi$^9$, P. Wiseman$^2$, R. Chen$^{10}$, E. Charleton$^2$, M. Acevedo$^8$, P. Armstrong$^{11}$, B.M. Boyd$^{12}$, 
    \newauthor
    D. Brout$^{13}$, R. Camilleri$^{7}$, J. Frieman$^{14}$, L. Galbany$^{15,16}$, M. Grayling$^{12}$, L. Kelsey$^{12}$, B. Rose$^{17}$, 
    \newauthor
    B. Sánchez$^{18}$,   
    J. Lee$^{19}$, A. Möller$^{20}$,  M. Smith$^{21}$, M. Sullivan$^{2}$, N. Shiamtanis$^{2}$, 
{A.~Alarcon}$^{22}$,
{S.S.~Allam}$^{14}$,
\newauthor
{F.~Andrade-Oliveira}$^{23}$, 
{S.~Avila}$^{24}$,
{D.~Bacon}$^{25}$,
{J.~Blazek}$^{26}$, 
{S.~Bocquet}$^{27}$, 
{D.~Brooks}$^{3}$,
{D.~L.~Burke}$^{28,29}$, 
\newauthor
{A.~Carnero~Rosell}$^{30,31}$, 
{J.~Carretero}$^{32}$, 
{R.~Cawthon}$^{33}$, 
{L.~N.~da Costa}$^{31}$,
{M.~E.~da Silva Pereira}$^{34}$, 
\newauthor
{H.~T.~Diehl}$^{14}$,
{S.~Dodelson}$^{5,14,6}$,
{P.~Doel}$^{3}$,
{S.~Everett}$^{35}$, 
{C. Frohmaier}$^{25}$,
{J.~Garc\'ia-Bellido}$^{36}$, 
{D.~Gruen}$^{27}$,
\newauthor
{G.~Gutierrez}$^{14}$,
{K.~Herner}$^{14}$,
{S.~R.~Hinton}$^{7}$,
{D.~L.~Hollowood}$^{37}$, 
{K.~Honscheid}$^{38,39}$, 
{D.~Huterer}$^{40}$, 
\newauthor
{D.~J.~James}$^{41}$, 
{N.~Jeffrey}$^{3}$,
{K.~Kuehn}$^{41,42}$, 
{O.~Lahav}$^{3}$,
{S.~Lee}$^{43}$, 
{C.~Lidman}$^{43,11}$, 
{J.~L.~Marshall}$^{44}$, 
\newauthor
{J. Mena-Fern{\'a}ndez}$^{18}$,
{F.~Menanteau}$^{45,46}$, 
{R.~Miquel}$^{47,48}$, 
{J.~Muir}$^{49,50}$, 
{J.~Myles}$^{51}$, 
{R.~L.~C.~Ogando}$^{52,53}$, 
\newauthor
{M.~Paterno}$^{14}$,
{A.~A.~Plazas~Malag\'on}$^{28,29}$,
{A.~Porredon}$^{24,54}$, 
{J.~Prat}$^{5,55}$, 
{R.C.~Nichol}$^{56}$, 
{A.~K.~Romer}$^{57}$, 
\newauthor
{A.~Roodman}$^{28,29}$,
{E.~Sanchez}$^{24}$,
{D.~Sanchez Cid}$^{24,23}$,
{I.~Sevilla-Noarbe}$^{24}$,
{E.~Suchyta}$^{58,45}$, 
\newauthor
{M.~E.~C.~Swanson}$^{45}$,
{C.~To}$^{5}$,
{D.~L.~Tucker}$^{14}$,
{A.~R.~Walker}$^{59}$, 
{N.~Weaverdyck}$^{60,61}$ 
and {M. Aguena}$^{62}$
}

\date{\centering The DES Collaboration}

\pubyear{2025}

\begin{document}
\label{firstpage}
\pagerange{\pageref{firstpage}--\pageref{lastpage}}

\maketitle


\begin{abstract}
We present improved cosmological constraints from a re-analysis of the Dark Energy Survey (DES) 5-year sample of Type Ia supernovae (DES-SN5YR). This re-analysis includes an improved photometric cross-calibration, recent white dwarf observations to cross-calibrate between DES and low redshift surveys, retraining the SALT3 light curve model and fixing a numerical approximation in the host galaxy colour law. Our fully recalibrated sample, which we call DES-Dovekie, comprises $\sim$1600 likely Type Ia SNe from DES and $\sim$200 low-redshift SNe from other surveys. With DES-Dovekie, we obtain $\Omega_{\rm m} = 0.330 \pm 0.015$ in Flat $\Lambda$CDM which changes $\Omega_{\rm m}$ by $-0.022$ compared to DES-SN5YR. Combining DES-Dovekie with CMB data from Planck, ACT and SPT and the DESI DR2 measurements in a Flat $w_0 w_a$CDM cosmology, we find $w_0 = -0.803 \pm 0.054$, $w_a = -0.72 \pm 0.21$. Our results hold a significance of $3.2\sigma$, reduced from $4.2\sigma$ for DES-SN5YR, to reject the null hypothesis that the data are compatible with the cosmological constant. This significance is equivalent to a Bayesian model preference odds of approximately 5:1 in favour of the Flat $w_0 w_a$CDM model. Using generally accepted thresholds for model preference, our updated data exhibits only a weak preference for evolving dark energy. 
\end{abstract}




\section{Introduction}\label{sec:Intro}

Type Ia supernovae by virtue of their brightness ($\sim -19$ mag at peak) and low scatter ($\sim 0.15$ mags after standardisation) provide precise distance measurements at gigaparsec scales. Used to discover the accelerating expansion of the universe in \cite{Riess98, Perlmutter99} with $\sim$50 SNe\,Ia, competitive modern analyses require thousands of SNe \citep{Rubin25, Popovic24a, Brout22}. In 2024, the Dark Energy Survey (DES) released \citep{Sanchez24} and analysed \citep{DES5YR, DES5YRKP} their 5-year sample of $\sim1600$ SNe, the largest single-telescope sample of likely SNe~Ia used in a cosmology analysis to date. Measurements of baryon acoustic oscillations (BAO), the cosmic microwave background (CMB), and gravitational lensing provide complementary constraints. Jointly, these probes provide impressive measurements on the components and history of the universe \citep{SDSSBAO, DES5YRBAO, DESBAO24,DESIDR2}. 

The concordance cosmological model $\Lambda$CDM, with the universe composed of a cosmological constant, dark matter, and baryonic matter, has been the \textit{de-facto} cosmological model for the past 25 years. An empirical way to parametrise deviations from $\Lambda$CDM is by the dark energy equation-of-state parameter, $w = P/\rho$, where a cosmological constant $\Lambda$ corresponds to $w=-1$. A conventional parametrisation for a time-evolving equation of state is $w(a) = w_0 + (1-a)w_a$, where $a=1/(1+z)$ is the scale factor \citep{chevallier2001, linder2003}. This parametrisation is capable of approximating the evolution of a wide variety of physical models \citep[eg][]{camilleri2024, Lodha2025} over the redshift range spanned by SNe. 

Until recently, analyses of cosmological datasets \citep{Betoule14, Scolnic18, DES3YR, Brout18SYS,Brout22} showed consistency with $\Lambda$CDM. However, recent supernova measurements from DES-SN5YR \citep{DES5YRKP} and Union3 \citep{Rubin25} show $\sim2.5\sigma$ evidence for time-varying dark energy, on their own or in combination with the CMB \citep{Planck20} and BAO from the Sloan Digital Sky Survey \citep{SDSSBOSS}\footnote{\url{https://www.sdss4.org/science/final-bao-and-rsd-measurements}}. Subsequently, new BAO measurements from the Dark Energy Spectroscopic Instrument (DESI, \citealp{DESIDR2}) revealed a $2.8-4.2\sigma$ discrepancy with a cosmological constant, depending on the choice of SN\,Ia sample. DESI results are confirmed by DES BAO\footnote{The DES BAO results use photometric redshifts, in contrast to the DESI spectroscopic redshifts.} which in combination with DES-SN5YR and the CMB show a $~3 \sigma$ discrepancy from $\Lambda$CDM \citep{DESBAO24,DES5YRBAO}. 

SN\, Ia data are vital to these results. While $\Lambda$CDM does not provide the best fit to the combination of BAO and CMB data in the absence of SNe Ia, models with constant dark energy but non-zero spatial curvature provide a reasonable alternative fit \citep{Chen2025}.

The consistency of the preference for evolving dark energy between DES-SN5YR and Union3 is particularly notable, due to the majority of the data being different and the pipelines being independent. Using the DES-SN5YR public data release, \citet{Efstathiou2024} noted that the preference for evolving dark energy relies on a very robust relative calibration to within 0.01 mag between SNe Ia across their full redshift range, involving multiple surveys. Additionally, \citet{Dhawan24} have demonstrated a toy model to illustrate how SN\, Ia systematics could potentially translate into a false preference for evolving dark energy. Accordingly, it is crucial to ensure SN Ia systematics are well understood and treated robustly. In particular, \citet{Efstathiou2024} noted differences averaging 0.04 mag between low-z and high-z SN\, Ia that are common to both the Pantheon+ and DES-SN5YR datasets. \citet{Vincenzi2025} traced these differences, finding that they {are} expected due to improvements in host galaxy modelling, light curve fitting (by the SALT method), and different corrections for selection bias between the two datasets. In particular, the photometrically-identified DES-SN5YR data set is more complete, requiring smaller bias corrections for selection than the spectroscopically-confirmed Pantheon+. As such, it may be reasonably argued that results from DES-SN5YR will have less systematic risk than those from the earlier Pantheon+. 

To further investigate the DES-SN5YR results, \cite{Dovekie} (hereafter, Dovekie) analysis focused on enhancing the photometric cross-calibration of SN Ia surveys, including the use of new data from the HST \citep{narayan2019,axelrod2023,Boyd25} and Gaia \citep{GAIAONE,GAIATWO,GAIATHREE}. Within the milieu of the current $\Lambda$CDM tension, and discussion of systematic uncertainties such as calibration, \citetalias{Dovekie} concluded that a reanalysis of DES-SN5YR with this new cross-calibration was necessary to properly estimate the impact on cosmological parameters. 

As part of the DES-SN5YR data release, the software and input files needed to replicate the analysis were included, with the aim of enabling future improvements and reanalyses. However, the data release was incomplete, missing a portion of the systematics. Here, we {realise the potential of} this release and complete it, conducting a full end-to-end reanalysis of the DES-SN5YR results using the \citetalias{Dovekie} calibration solution. This reanalysis includes: 1) verifying the pipeline by reproducing the results of \citetalias{DES5YR} and \cite{DES5YRKP}; 2) retraining the SALT3 model with Dovekie calibration; and 3) re-running the analysis pipeline (e.g., Fig 1 in \citetalias{DES5YR}) that includes light curve fitting, photometric classification, bias corrections, construction of the systematic covariance matrix, and cosmology fitting. We have also improved and simplified the interface to the publicly available DES-SN5YR pipeline to encourage future efforts to re-analyse this sample. 

After initially reproducing the original \citetalias{DES5YR} work, we address an outdated approximation in the implementation of the assumed dust colour law in \citetalias{DES5YR} (and Amalgame and Pantheon+) and assess its impact before performing our cosmological analysis with our improved calibration and fixed colour law. 

\begin{table}
    \centering
    \begin{tabular}{l|cc}
        Change & DES-SN5YR & DES-Dovekie \\
        \hline
         Calibration & \citetalias{fragilistic} & \citetalias{Dovekie} \\
         SALT Model & SALT3.DES5YR & SALT3.DOV \\
         Cal. Uncertainties & Underweighted & Fixed \\
         F99 Colour Law & Approximate & Exact \\
         Simulations & Generated & Regenerated \\
         & with SALT3.DES5YR & with SALT3.DOV \\
         BAO data & SDSS & DESI DR2 \\
         CMB data & Planck & Planck, ACT-DR6, SPT-3G \\
         Posterior Sampler & {MCMC} & {Nautilus} \\
    \end{tabular}
    \caption{A top-level summary of the changes between this work and \citetalias{DES5YR} and \citet{DES5YRKP}. Section \ref{sec:Differences} goes into more detail on each of these points.}
    \label{tab:toplevel}
\end{table}

Table \ref{tab:toplevel} provides a top-level summary of the changes we make to \citetalias{DES5YR} and \citet{DES5YRKP} pipelines, which constitute the `DES-Dovekie' result. The layout of the paper is as follows. Section \ref{sec:Data} provides a brief description of the DES-SN5YR data set, Section \ref{sec:Methodology} describes SN\,Ia standardisation and distance measurements. Section \ref{sec:CosmoMethods} describes cosmology fitting. We choose our range of models to facilitate comparisons with the literature, but discuss alternate parameterisations in this section. Differences between this work and DES-SN5YR are given in Section \ref{sec:Differences}, and a review of our consistency checks is presented in Section \ref{sec:Consistency}. This is followed by a description of our systematic uncertainties in Section \ref{sec:systdesc}, our Hubble Diagram in Section \ref{sec:Results}, and systematic uncertainties in Section \ref{sec:SYST}. Our cosmological results are presented in Section \ref{sec:Cosmo}, and we present our conclusions in Section \ref{sec:Conclusions}.

\section{Data}\label{sec:Data}

The data we use here is the same as used in the \citetalias{DES5YR} analysis, combining high-$z$ measurements of photometrically classified SNe with low-$z$ spectroscopically confirmed samples. All surveys involved have been recalibrated in the \citetalias{Dovekie} analysis. 

\subsection{DES-SN5YR}

The DES-SN program ran for 5 years using the Dark Energy Camera (DECam, \citealp{Flaugher15,DESCAM}), across ten 3 deg$^2$ fields; two deep fields extending to 24.5 mag depth per visit in each of the $griz$ bands, and eight shallow fields with 23.5 mag depth. Host-galaxy redshifts were acquired with a dedicated followup program OzDES on the AAOmega spectrograph \citep{Smith04,Lidman20}. Further details on the specific search strategy and spectroscopic follow-up programs are available in \cite{Kessler15} and \cite{Smith20}.

The published DES lightcurves are measured with the ``Scene Modelling Photometry'' pipeline (SMP, \citealp{Brout19b, Sanchez24}), simultaneously modelling the SN Ia and host-galaxy fluxes. The DES fluxes include chromatic corrections to account for SED differences between SNe\,Ia and calibration stars \citep{Lasker2019} and atmospheric corrections to account for Differential Chromatic Refraction \citep{LeeAcevedo23}. For the low-redshift samples, we did not update their published photometry.

This work, alongside numerous other fruits of the DES collaboration, resulted in the data and cosmology releases as seen in \cite{Sanchez24, DES5YR, Moller22}. We begin with the same initial sample of $\sim3600$ observed transients that were published in \citet{Sanchez24}, but due to survey cuts our re-calibrated dataset will not be the same size as DES-SN5YR. 

\subsection{Low-redshift}

The DES-SN5YR analysis eschewed some of the older historical samples of SNe Ia at low redshift; instead, the low-redshift anchor is comprised of CfA3 \citep{Hicken09a}, CfA4, \citep{Hicken12}, CSP \citep{Krisciunas17} (comprising the `Low-z' sample), and Foundation \citep{Foley17}. Only SNe Ia above $z = 0.025$ were included, mitigating the impact of peculiar velocities, and an additional 1\% mag error floor was added to the Foundation sample.  In contrast to the photometrically-typed SNe in DES, the low redshift supernovae are all spectroscopically confirmed. 


\subsection{Host Galaxies}

The most likely host galaxy for an SN Ia (and therefore the corresponding redshift) is identified in deep-stacked photometry free of SN Ia light \citep{Wiseman20} with the Directional Light Radius ($d_{\rm DLR}$) method presented by \cite{Sullivan06, Gupta16}. An SN Ia is considered \say{hostless} if the $d_{\rm DLR} >4$, otherwise likely hosts were targeted as part of the previously-mentioned OzDES programme \citep{Yuan15,Childress17,Lidman20}. The efficiencies of these followup programs in acquiring a host galaxy redshift are detailed in \cite{Vincenzi21} and \cite{Sanchez24}; SNe without a host are not included in our cosmology sample.

The host galaxies are characterised via two global properties: the stellar mass $M_{\star}$ and the rest-frame $u-r$ colour. These two properties can be computed across the redshift range of the analysis with the limited broadband photometry. The host galaxy masses and colours are fit consistently across the surveys using the code, pipeline, and procedure described in \cite{Sullivan10} and the PEGASE2 templates from \cite{Fioc99,LeBorgne02} with a \cite{Kroupa01} initial mass function. The DES host galaxies are supplemented with photometry from $uJHK$ photometry when possible \citep{Wiseman20, Hartley22}, and the low-redshift host photometry from PS1 is supplemented with UV photometry from GALEX \cite{Bianchi17} and the SDSS $u-$band.

For the low-redshift anchor, updated spectroscopic redshifts are provided by \cite{Carr21}; estimated peculiar velocity corrections are provided by \cite{Peterson21} with uncertainties set to a uniform 250 km s$^{-1}$ \citep{Scolnic18}.

\subsection{Non-Ia Classification}

Baseline classification of SNe in DES-SN5YR is performed with SuperNNova ({SNN}, \citealp{Moller19}),  a recurrent neural network machine learning classifier designed to train and operate on photometric SNIa data and host-galaxy redshift. To train SuperNNova, a suite of realistic simulations is generated using SED time-series for core collapse (SN types II, Ib, Ic) and for peculiar 91bg-like and SNIax \citep{Vincenzi19, Kessler19}, with SNN model parameters optimised for DES in \cite{Moller24}.

Alternative classifiers SCONE \citep{Qu21} and SNIRF\footnote{https://github.com/evevkovacs/ML-SN-Classifier}, each with different training and prediction algorithms compared to SNN, are used  to assess systematic uncertainties on classification (for more details, see \citealp{DES5YR, Moller22}). 

\section{Standardisation Methodology}\label{sec:Methodology}

Here we provide a brief summary of the cosmological inference methodology employed by the DES-SN5YR analysis. Broadly, The BEAMS with Bias Correction (BBC) method \citep{Kessler16} (implemented within \cite{SNANA, Kessler19} and managed by \citealp{PIPPIN}) uses the light curve parameters to determine distances, and to apply bias corrections that account for selection effects and non-SNIa contamination. In the cosmology fit, these distances are weighted by their uncertainties and their probability of being a type Ia supernova.

\subsection{Lightcurve Modelling and Fitting}

To acquire SN~Ia distances, we perform multi-band photometry fitting of the observed photometry of the SN Ia light curve. \citetalias{DES5YR} used the SALT3 model \citep{Kenworthy21}, itself an update to the widely-used SALT2 model from \cite{Guy10}. SALT is a description of the spectral energy distribution (SED) of the population of observed type Ia supernovae, similar to a principal component analysis; it is empirically derived from a training sample of spectroscopically confirmed SNe. \citetalias{DES5YR} used the \citet{Taylor23} training of SALT3 (hereafter SALT3.DES5YR), which presented four major changes to the previous models used in cosmology: an increased training sample, increased wavelength coverage\footnote{$2800-8000$ \r{A} central filter wavelength compared to $2800-7000$ \r{A}.}, the eschewing of observer-frame $U$ band from the training, and the Fragilistic calibration solution from \cite{fragilistic}. Here, we use the updated SALT3.DOV model, retrained using the fiducial \citetalias{Dovekie} calibration solution. Further, the SALT3.DOV model has cut many SNe measured by legacy surveys from the training sample; the SALT3.DOV model thus continues a trend towards more stringent requirements on calibration.

For a given light-curve fit, there are five parameters used to characterize the SN: redshift $z$, time of peak brightness $T_0$, stretch $x_1$, colour $c$, and the overall flux normalization $x_0$, typically used in magnitude units $m_B$ ($m_B = 10.635 -2.5\log(x_0)$). The best-fit values and associated uncertainties of each parameter are determined to measure distances; to this end, spectroscopic measurements of the host-galaxy redshift are used ($\sigma_z \sim 10^{-4}$), allowing us to fix the redshift during the lightcurve fitting process.  

Distances with the SALT model are inferred from the Tripp estimator \citep{Tripp98}
\begin{equation}\label{eq:Tripp}
    \mu = m_B +\alpha x_1 -\beta c - M_0 - \Delta\mu_{\rm bias} - \delta_{\rm host},
\end{equation}
where $m_B$, $x_1$, and $c$ are defined as above, $M_0$ is the peak brightness of a fiducial SN Ia.
$\delta_{\rm host}$ is the term that corrects for the `mass step' \citep{Kelly10, Sullivan10, Lampeitl10} that occurs for the standardised brightness of an SN Ia, {expressed as a sigmoid function}:

\begin{equation}
    \delta_{\rm host} = \gamma (1+e^{(M_{\star} - S)/\tau_{M_{\star}}} )^{-1} - \gamma/2,
\end{equation}
where $\gamma$ is the magnitude of the luminosity difference between those SNe located in `high' mass galaxies $(M_{\star} > 10^{10} M_{\odot})$ and `low mass' $(M_{\star} < 10^{10} M_{\odot})$ galaxies. $S$ is the step location, nominally set to $S = 10^{10}M_{\odot}$, and $\tau_{M_{\star}}$ is the width of the step. The mass step is accounted for during the BBC process.
$\alpha$, $\beta$ and $\Delta\mu_{\rm bias}$  are explained below. 

\subsection{Bias Corrections}

The $\alpha$ and $\beta$ nuisance\footnote{{Not `nuisance' in the Bayesian sense, as we do not marginalise these parameters.}} parameters describe the luminosity-stretch and luminosity-colour relationship respectively. These nuisance parameters, alongside the luminosity-distance relationship of SNe Ia, {can be biased} by selection effects arising from the flux-limited nature of SNe Ia observations. The biases arising from these selection effects must be accounted for, and typically this is done via large Monte Carlo simulations that attempt to accurately model detection and other selection effects \cite{ Perrett10, Betoule14, Kessler19, Popovic21a}. 

Following DES-SN5YR, we use the BBC update in \cite{Popovic22} that is compatible with dust models introduced in \citet{BS20}, and we use the updated dust parameters from  \cite{Popovic22}. The simulated $\Delta \mu_{\rm bias}$ term is averaged in 4D cells of $\{z,c,x_1,\rm{log}_{10}(M_{\star})\}$:
\begin{equation}\label{eq:biascor}
    \Delta \mu_{\rm bias} = m_B + \alpha^{\rm true} x_{1} - \beta^{\rm true} c - M_0^{\rm true} - \mu^{\rm true},
\end{equation}
where parameters noted with the superscript `true' are the simulated parameters, and $m_B$, $x_1$, and $c$ are obtained from SALT3.Dovekie light curve fits in the same was as for the data. 

\subsection{Non-Ia Contamination}

In a cosmology analysis with photometrically-classified SNe Ia, a portion of SNe that remain after light curve fitting quality cuts may not actually be Type Ia. In order to properly account for these contaminants in the cosmology analysis, DES-SN5YR made use of the Bayesian Estimation Applied to Multiple Species (BEAMS) framework from \cite{Knights13, Kunz07, Hlozek12} and implemented in BBC. {See \cite{Vincenzi21, DES5YR} for more information.}


\section{Cosmology Fitting}\label{sec:CosmoMethods}

\subsection{Models and Theory}

Our baseline model for cosmological distances is the linear parameterisation of the dark energy equation of state, given by 
\begin{equation}
    w=w_0+w_a(1-a) \;,
\end{equation}
and denoted $w_0w_a$CDM. We opt for this parameterisation (although suboptimal, as $w_0$ and $w_a$ will in general be correlated, this formulation has now become conventional). We define the curvature density parameter $\Omega_{\rm k}$ as $\Omega_{\rm m} + \Omega_\Lambda = 1 - \Omega_{\rm k}$, and refer to $\Omega_{\rm k}=0$ as flat. We omit the radiation density as negligible for late-times. Then, Flat $\Lambda$CDM is a special case of Flat $w_0w_a$CDM with $w_0 = -1$ and $w_a = 0$, and Flat $w$CDM means $w_a = 0$ but $w_0$ is allowed to vary from $-1$. We collectively denote the cosmological model parameters as $\Theta = (\Omega_{\rm m}, \Omega_\Lambda, w_0, w_a)$ where $\Omega_{\rm m}$ is the matter density parameter today.

A potentially more informative description of the equation of state is given by $(w_p, w_a)$ where $$w=w_p+w_a(a_p-a)$$ and $a_p$ is defined as the scale factor where the variance of $w(a)$ is minimized, or equivalently where $w_p$ and $w_a$ are uncorrelated. As explained in \citet{Linder2007}, it is expected to be the case that when including CMB data that $a_p \sim 0.4$, nevertheless this parametrisation may be interesting for some data combinations.

The transverse comoving distance $\chi$ is
\begin{equation}
    \chi(z_{\rm cos}) = \frac{c}{H_0} \frac{1}{\sqrt{|\Omega_{\rm k}|}} \;\rm sinn \big(\sqrt{|\Omega_{\rm k}|}\int_0^{z_{\rm cos}}\frac{dz}{E(z)}\big) \;\;,
\end{equation}
where $z_{\rm CMB}$ is the redshift in the rest-frame of the CMB (the frame in which the dipole would be absent), and sinn$(x)$ = $\sin(x), x, \sinh(x)$ depending on $\;\Omega_{\rm k}<0, \Omega_{\rm k}=0, \Omega_{\rm k}>0$ respectively. $E(z)\equiv H(z)/H_0$ is the normalized redshift-dependent expansion rate and 
\begin{equation}
\begin{split}
    H(z) = \big[ \Omega_{\rm m} (1+z)^3 & +\Omega_{\rm k} (1+z)^2+ \\
    & \Omega_\Lambda (1+z)^{3(1+w_0+w_a)}e^{-3w_a z/(1+z)} \big]^{1/2} \;\; .
\end{split}
\end{equation}
The luminosity distance is given by
\begin{equation}
D_l (z_{\rm obs},z_{\rm CMB}) = (1+z_{\rm obs})\chi(z_{\rm CMB}) \;\;, 
\end{equation}
where $z_{\rm obs}$ is the observed heliocentric redshift, which captures the effect of beaming due to peculiar velocity of both the Sun and the SN Ia with respect to the CMB rest frame. The distance modulus is $\mu(z,\Theta) = 5 \log_{10}(D_l (z,\Theta)/10~{\rm Mpc}) + 25$.

We compute the difference between data and theory for every $i$th supernova, $\Delta \mu_i = \mu_{{\rm obs},i} - \mu(z_i,\Theta)$, and write the likelihood $\mathcal{L}$ in the standard form 
\begin{equation} 
-2 \log{\mathcal{L}} \equiv \chi^2 = 
\Delta \mu_i \mathcal{C}_{ij}^{-1} \Delta \mu_j^T \;\;,
\end{equation}
where $\mathcal{C}^{-1}$ is the inverse covariance matrix (including both statistical and systematic errors) of the $\Delta \mu$ vector, and the methodology of computing $\cal C_{\rm syst}$ follows the approach described in \citet{DES5YR}:
\begin{equation}\label{eq:covsyst}
    \mathcal{C}^{ij}_{\rm syst} = \sum^{N_{\rm syst}}_{S=1} (\Delta\mu^i_{\rm obs,S})(\Delta\mu^j_{\rm obs,S})W^2_S
\end{equation}
where changing systematic parameter $S$ gives differences in SN Ia distances $\Delta\mu$. Indices $i,j$ are iterated over all the SNe in the analysis ($i,j = 1 ... N_{\rm SNe}$) and $W_S$ is the weight for each systematic (which is set to $1$ unless otherwise stated). 

A potential source of confusion in cosmology with SN Ia is that the absolute magnitude of the fiducial SN\,Ia ($M_0$) and the $H_0$ parameter (which appears in the luminosity distance) are completely degenerate. They may be combined in the single parameter $\mathcal{M}=M_0+5\log_{10}(c/H_0)$. Our results are marginalised over $\mathcal{M}$.

For our nominal cosmology, we use the {Cosmosis}\footnote{\url{https://github.com/joezuntz/cosmosis}} package from \cite{zuntz} with the {Nautilus}\footnote{\url{https://github.com/johannesulf/nautilus}} nested sampler \citep{nautilus}.

\subsection{Combination with other cosmology probes}

We combine our DES SNe constraints with those from other complementary probes. 

\begin{itemize}
    \item \textbf{Cosmic Microwave Background:} We use the measurements of temperature and polarisation power spectra (TTTEEE) from \cite{Plik2020}, in combination with ground-based measurements from the Atacama Cosmology Telescope (ACT) and the South Pole Telescope (SPT). We also include lensing reconstructions in our analysis. Specifically, we combine the \texttt{simall} and \texttt{Commander} likelihoods for $\ell < 30$, with the \texttt{Plik-lite} likelihood for $\ell < 1000, 600$ (TT, and TE, EE respectively) as wrapped in the Python implementation \texttt{Planck-py}\footnote{\url{https://github.com/heatherprince/planck-lite-py}} \citep{PrinceDunkley19}. To this we add TTTEEE data implemented in the \texttt{ACT-DR6-Lite}\footnote{\url{https://github.com/ACTCollaboration/DR6-ACT-lite}} likelihood \citep{ActDr6maps, ActDr6like}, and SPT-3G data \citep{Camphuis2025} as wrapped in the \texttt{candl} likelihood \citep{Balkenhol2024}\footnote{\url{https://github.com/Lbalkenhol/candl}}. Our lensing likelihoods are the \texttt{actplanck-baseline}\footnote{\url{https://github.com/ACTCollaboration/act_dr6_lenslike}} option based on the combined ACT and Planck lensing reconstruction maps \citep{Madhavacheril2024, Qu2024, Carron2022}, together with SPT-3G lensing \citep{Pan2023} again implemented in \texttt{candl}. This forms a comprehensive and up-to-date account of CMB data, and it extends that used in both \citet{DES5YRKP} and \cite{DESIDR2} by including both ACT and SPT. While there is a small $\ell$ overlap between ACT and Planck, and a small sky area overlap between ACT and SPT {that may cause their constraints to not be fully independent}, it has been argued that the overlap is not material to cosmological analysis \citep{ActDr6like, Camphuis2025}. We comment on the influence of the choice of CMB data further in the results section.
    \item \textbf{Baryonic Acoustic Oscillations:} We use data from DESI DR2 as presented in \citet{DESIDR2}. DESI DR2 measures the apparent size of baryon acoustic oscillations (BAO) both along and perpendicular to the line of sight of various tracers in redshift bins ranging from $0.3 < z< 2.3$. These act to constrain expansion models between the distance between the CMB and the relevant redshift.
\end{itemize}

\subsection{Model preference}

To interpret our parameter constraints, we test the relative preference of extended models for our full range of data compared to Flat $\Lambda$CDM. 

There has been some debate in the community as to appropriate metrics for model preference, with some authors preferring frequentist methods \citep[e.g.][]{DESIDR2}, and others preferring Bayesian evidence \citep[e.g.][]{DES5YRKP}. For both methodologies, preference is primarily driven from the goodness-of-fit improvement $\Delta \chi^2$. The difference between the two can then be largely attributed to the penalty applied to the complex model, or equivalently appropriate thresholds against which to judge significance.

Frequentist methods evaluate relative probabilities of data, not models. As per \citet{DESIDR2}, we apply Wilk's Theorem\footnote{Strictly only approximate for real data, as terms of $\mathcal{O}(1/\sqrt{N})$ where $N$ is the number of data points are neglected.} \citep{wilks1938}. This states that the logarithm of the ratio of the maximum likelihood (ML) probability of the same data in each model follows a $\chi^2$-distribution with degrees of freedom (d.o.f.) equal to the number of additional parameters in the extended model (one in the case of $w$CDM and two for $w_0 w_a$CDM). We determine the ML for each model and the difference $\Delta \chi^2_{\rm ML}$ between the extended model and Flat $\Lambda$CDM (by construction a positive value), and convert to a $p$-value with the $\chi^2$-distribution. We then express the $p$-value as a number of sigma, $n\,\sigma$, by solving
\begin{equation}
\label{eq:wilk}
\rm CDF_{\Delta \chi^2} (\log p \; | \; \rm d.o.f.) = \frac{1}{\sqrt{2\pi}} \int_{-n}^{n} e^{-t^2/2} dt \;\;.
\end{equation}
In this case, the complex model penalty is relatively lenient, appearing as the degrees of freedom in the $\chi^2$-distribution.

Bayesian methods consider directly the relative confidence of models, given the data. The evidence ratio $R_{01}$ is defined as 
\begin{equation}
\label{eq:bayesian}
R_{01}  = \frac{p(M_0 | D)}{p(M_1 | D)} \;\;,
\end{equation}
where $M_0$ is Flat $\Lambda$CDM and $M_1$ is the extended model and $D$ is the data. Given an agnostic a-priori preference for either model, $\log R_{01} = \Delta \log \mathcal{Z}$ where the evidence $\mathcal{Z}$ of each model is 
\begin{equation}
\label{eq:evidence}
\mathcal{Z} = \int p(D | \Theta, M) p(\Theta | M) d\Theta \;\;,
\end{equation}
which has explicit dependence on the a priori confidence in the model parameters $p(\Theta)$. In this case, the complex model penalty is approximated by the relative compression between the a priori and a posteriori probability volumes \citep[see e.g.][]{SiviaBook}. $\log \mathcal{Z}$ is computed by {Nautilus} as part of the sampling process. Although {Nautilus} does not compute errors in this statistic, we have tested alternative samplers, {Nautilus} settings and choices of random seed for initialisation, and estimate our error to be $\sigma (\log \mathcal{Z}) \sim 0.2$.  We interpret $\Delta \log \mathcal{Z}$ on the scale defined in Table I of \citet{Trotta2008}, as used by \citet{DES5YRKP}. 

The arbitrariness of Bayesian priors has been sometimes given as a reason to prefer frequentist methods. However, reasonable prior confidence intervals can be set based on broad astrophysical considerations; for example, a minimum age for the Universe based on stellar ages. If there is a decisive model preference, the widths of the priors (so far as they remain reasonable) will not be a determining factor. We quote both metrics.

For the Bayesian evidence calculations, we adopt uniform priors $H_0 \in (0.55,0.91)$, $w_0 \in U(-3, -0.4)$, $w_a \in U(-3,2)$, $\Omega_{\rm k} \in (-0.15, 0.15)$ and $\Omega_{\rm m} \in (0.1, 0.5)$. As is conventional, we also require $w_0+w_a < 0$.\footnote{$w_0 + w_a < 0$ is justified in the literature as necessary to ensure there is a period of matter-domination in the early universe (necessary for growth of structure).} For the CMB, we adopt $\Omega_{\rm b} \in (0.03,0.07)$, $\tau \in (0.01,0.2)$, $10^9 A_s \in (0.5,5.0)$ and we fix the sum of neutrino masses at $\Sigma m_{\nu} = 0.06$ eV. The CMB normalisation nuisance parameter priors are $A_{\rm Planck} \sim \mathcal{N}(1.0,0.0025)$, $P_{\rm ACT} \sim \mathcal{N}(1.0,0.003)$, $E_{\rm cal} \sim \mathcal{N}(1.0,0.0095)$, $T_{\rm cal} \sim \mathcal{N}(1.0,0.0036)$, $A_{\rm foreground} \in (0.0,2.0)$ \citep[as recommended in][]{Camphuis2025} and we fix the remaining ACT normalisation parameter $A_{\rm ACT}$ to $P_{\rm ACT}$ as ACT is calibrated off Planck.

While our priors are broadly consistent with \cite{DESIDR2}, there are a few differences. Firstly, we have lowered the upper limit on $\Omega_{\rm m}$ from 0.9 to 0.5 because $ \Omega_{\rm m} >0.5$ is inconsistent with a wide range of other galaxy surveys. Secondly, the upper limit on $w_0$ is set to avoid confusing dark energy with other components of the energy density such as curvature or matter. Thirdly, the prior on $\Omega_{\rm k}$ is tighter than typically adopted. This more restrictive prior reduces compute time to acceptable levels, and is consistent with the evidence to date in favour of a Universe that is close to spatially flat. Our posteriors are wholly contained within priors, except for some SN-only cases. In this instance, we quote parameter constraints on broadened priors, while retaining the above priors for the evidence calculations.

\section{Differences to DES-SN5YR}\label{sec:Differences}

In the introduction, we briefly mentioned a mistake in the implementation of the \cite{Fitzpatrick99} colour law. {SNANA} was using a polynomial expansion centred at $R_V=3.1$, which was accurate to $<1\%$ for this assumed $R_V$ value but for simulations of bias corrections where a multitude of $R_V$ are used, the error is larger. This mistake was identified in the process of other work within {SNANA}, and replaced with the full \cite{Fitzpatrick99} colour law. This change has negligible impact on the data with the assumed $R_V=3.1$ for Milky Way extinction. There is a measurable impact on simulated bias corrections as described in Appendix \ref{sec:F99}.

\citetalias{Dovekie} performed a re-calibration of historically-used samples of SNe Ia, including the low-redshift surveys used within the DES-SN5YR analysis. While \citetalias{Dovekie} uses the same overall framework as the DES-SN5YR calibration solution, (Fragilistic; \citealp{fragilistic}) using the all-sky Pan-STARRs (PS1) telescope\footnote{DES does not have sufficient sky-coverage overlap with low-redshift surveys, which precludes its use here as the interstitial.} as interstitial observations, \citetalias{Dovekie} improves on this methodology in a number of ways:
\begin{itemize}
    \item The addition of DA white dwarfs from \cite{Boyd25}, combining extensive modelling of these white dwarf spectra with direct observations from DES, PS1, and the Sloan Digital Sky Survey (SDSS).
    \item The use of Gaia spectroscopy as a complementary method to characterise published filters by integrating spectra through filter transmission functions, alongside filter transmission measurements with PS1.
    \item Subtle improvements to error modelling in the cross-calibration. 
    \item Separate improvements to the SALT training and error models. 
\end{itemize}

In addition to a changed set of calibration offsets, \citetalias{Dovekie} solved for filter shifts, finding changes to the following filters that directly impact the DES-SN5YR cosmology analysis: CfA3K-$V$, CSP-$B/V$, and every filter in CfA4P1/2. The +30 \AA shift to the PS1 $g$ band in \citetalias{fragilistic} was reverted to its published \cite{Tonry12} value. 
These new filter changes, from fits to survey photometry, alongside the new calibration offsets from Dovekie, were used to train a new SALT model.

\begin{figure}
    \centering
    \includegraphics[width=9cm]{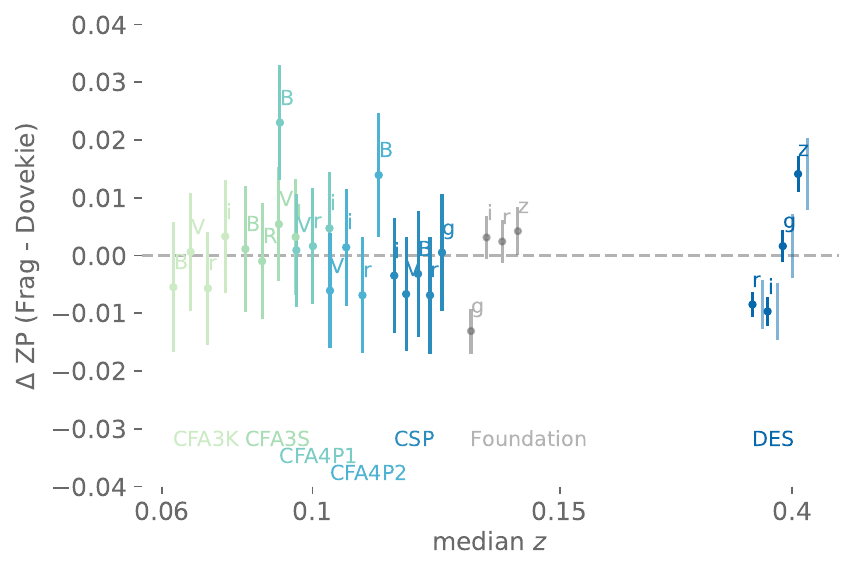}
    \caption{The DES-SN5YR sample includes the DES, CfA3S, CfA3K, CSP, and Foundation; the difference in calibration offsets for each filter and each survey is plotted at the approximate median redshift of the survey. To show the improvement in precision with the new calibration, the light-blue vertical bars show the calibration uncertainties for DES without using the nominal DA WD. The mean zero point across all surveys has been subtracted out for both Dovekie and Fragilistic, for visual clarity.}
    \label{fig:deltazp}
\end{figure}

These improvements, alongside the other system changes, have resulted in changes to the zero points of surveys within the DES-SN5YR sample. We show the changes, as a function of the median redshift of the survey, in Figure \ref{fig:deltazp}. A more thorough overview of this new calibration, and potential changes, is presented in \citetalias{Dovekie}. 

We repeat the DES-SN5YR cosmology analysis with this new calibration solution and SALT model, SALT3.DOV. We regenerate the `biasCor' files (Section \ref{sec:Methodology}) with the new SALT model, filters, and calibration offsets. Furthermore, we redo the DES-SN5YR light curve fits with this new SALT model and calibration. 

Additionally, during the course of this reanalysis, we found a minor error in the weighting (Equation 11 in \citetalias{DES5YR}) of the photometric uncertainty systematics. This error arose from a mistake in the input file,\footnote{Specifically, the weight on each of the 9 SALT3 calibration variants was mistakenly rounded to 0.3 instead of the correct value of $1/\sqrt{9} = 0.33$.} rather than an issue with the equation or methodology, and caused the weights of the calibration systematic to sum to 0.81 instead of 1. This resulted in a reduction of the estimated total photometric uncertainty by $\sim20$\% for \citetalias{DES5YR}. We correct this mistake.

Finally, we modify the cut on $\sigma_{x_{1}} < 1.0$ from \citetalias{DES5YR} to  $\sigma_{x_{1}} < 1.15$. Due to changes in error propagation between SALT3.DES5YR and the new SALT3.DOVEKIE, the $x_1$ errors were generally increased relative to other light curve fitting parameters. We choose the new $\sigma_{x_{1}} < 1.15$ cut to preserve the 20\% quantile cut established in \citetalias{DES5YR}.

To summarise:
\begin{itemize}
    \item \textbf{Updated \cite{Fitzpatrick99} Colour Law:} During unrelated work within SNANA,\footnote{See Appendix \ref{sec:F99}.} it was discovered that the implementation of the \cite{Fitzpatrick99} colour law term within SNANA was a polynomial expansion centered at $R_V = 3.1$. The full F99 colour law was implemented, and it was discovered that this impacted simulated and inferred distance moduli on the order of $\sim0.01$ mag, though primarily impacting the simulated bias corrections.  
    \item \textbf{Updated Calibration:} We use the updated cross-calibration solution from \citetalias{Dovekie}, in place of \citetalias{fragilistic}. While the ZP offsets between the two are largely similar, they notably differ for DES.
    \item \textbf{Updated SALT model:} \citetalias{fragilistic} combined the estimation of calibration uncertainties with those of the SALT modelling, which provided more reliable estimations of systematic uncertainty arising from the two. This necessitates a retraining of the SALT model with the new calibration solution. 
    \item \textbf{Re-generated Bias Correction Simulations:} Table \ref{tab:syst_description} highlights the specific systematic uncertainties that are regenerated, and therefore directly impacted, by this reanalysis. 
    \item \textbf{Fixed Calibration Uncertainties:} As previously stated, \citetalias{DES5YR} used an incorrect weight of $\psi = 0.3$ for each of the systematic SALT surfaces. This caused \citetalias{DES5YR} to underestimate the photometric uncertainty by $\sim20\%$. We update this to the more correct value of $0.33$.
\end{itemize}

\section{Consistency Checks}\label{sec:Consistency}

\begin{figure}
    \centering
    \includegraphics[width=9cm]{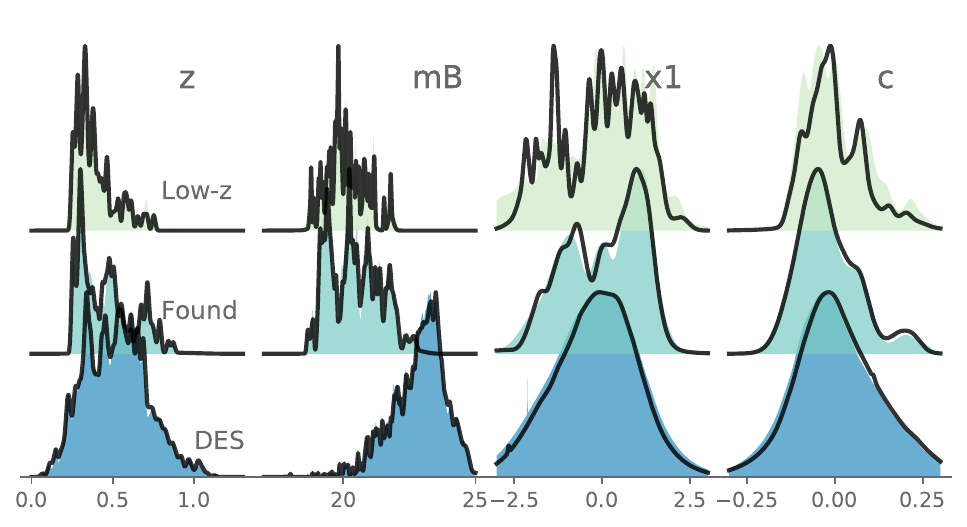}
    \caption{Distributions of $z, m_B, x_1,c$ for \citetalias{DES5YR} (black) and this work (filled histogram). We split the sample into its constituent parts, DES (blue), Foundation (teal), and Low-z (light green). We see agreement between the two analyses, with the exception of the DES $c$ distribution.}
    \label{fig:deltadist}
\end{figure}

In this section, we review the changes in SN Ia parameters between this work and the original DES-SN5YR analysis, and investigate the source of any changes and their potential impact. 

\begin{figure*}
    \centering
    \subfloat[\centering Hubble diagram of DES-Dovekie. DES is shown in blue, with low-redshift supernovae in orange. For each DES event, the classification probability from {SNN} is indicated by the colour. \textbf{Upper:} Full Hubble diagram.  \textbf{Lower:} Hubble residuals from best-fit $w_ow_a$CDM cosmology. The \cite{Planck18} cosmology is shown in light maroon dashed line. \textbf{Inset:} Redshift histogram of the two subsamples.  ]{{\includegraphics[width=18cm]{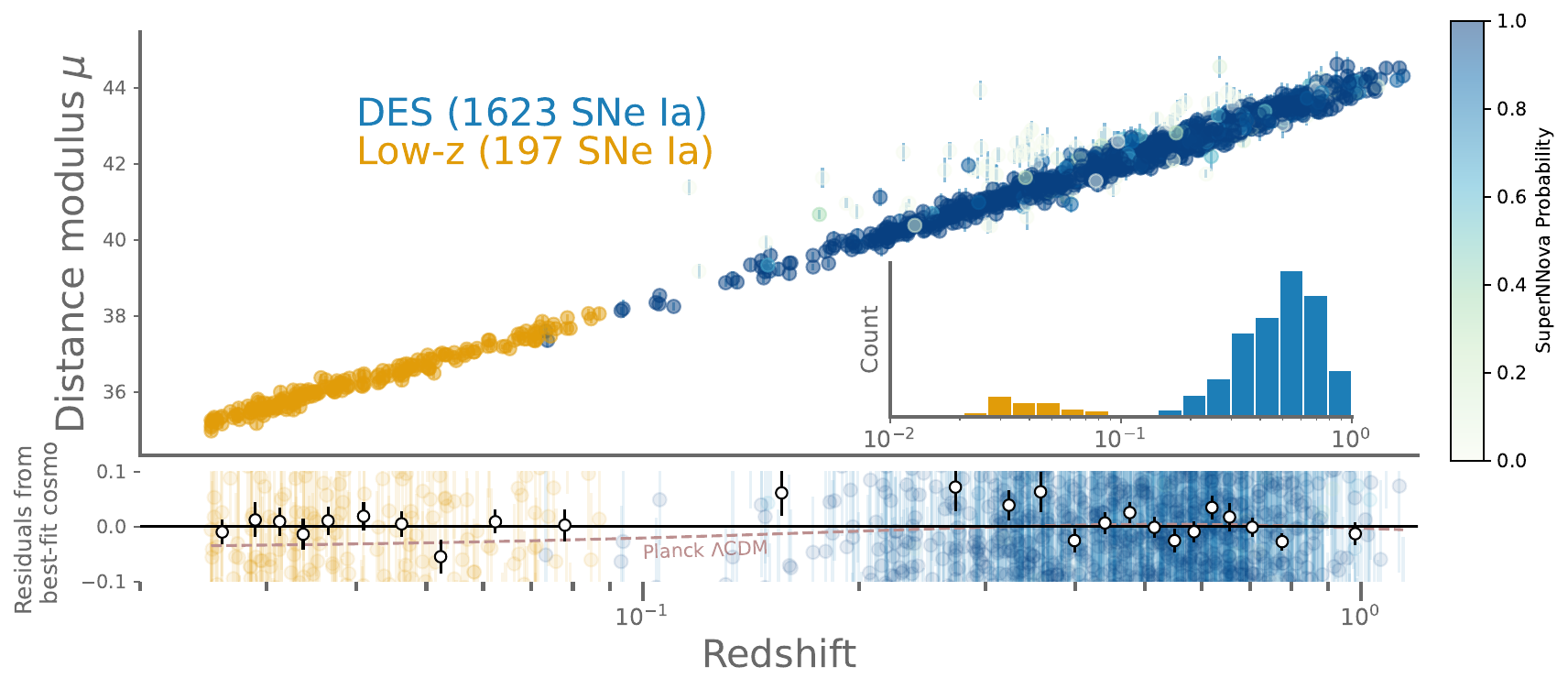}}}
    \qquad
    \subfloat[\centering The difference between the \citetalias{DES5YR} published distances and DES-Dovekie distances. The error-weighted average $\Delta \mu$ in black includes shift in $\mu$ {\em and} the change in bias-correction. We show the un-bias-corrected $\mu$ averages in blue, and the bias-correction component alone in orange. ]{{\includegraphics[width=18cm]{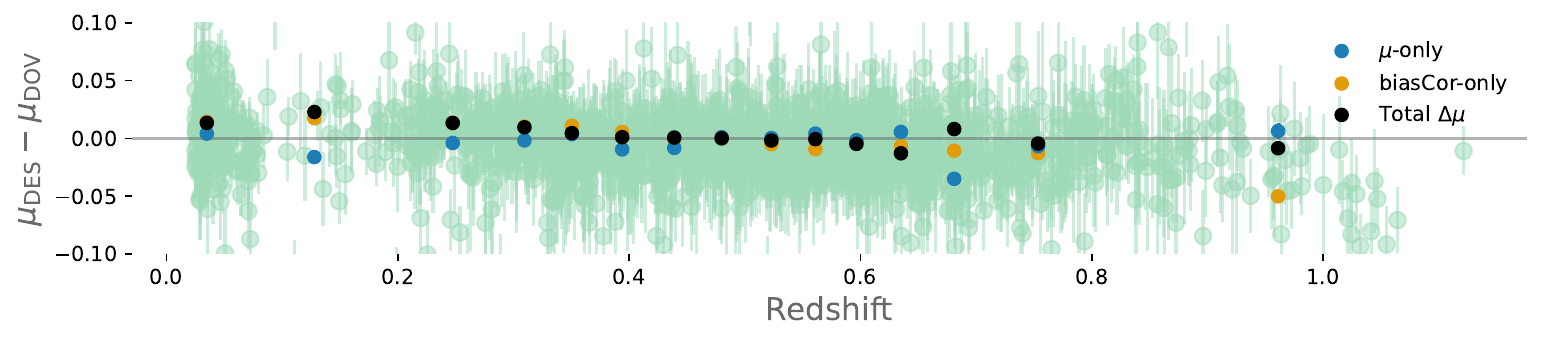} }}%
    \caption{The DES-Dovekie Hubble Diagram, and the change in inferred, bias-corrected $\mu$ values between \citetalias{DES5YR} and this work. }\label{fig:HUBBLE}
\end{figure*}

\subsection{\citetalias{DES5YR} and this work}

Figure \ref{fig:deltadist} shows the $z$, $m_B$, $x_1$, and $c$ distributions for the DES-SN5YR data set, fit with SALT3.DES5YR and Dovekie. \citetalias{DES5YR} contained 1829 SNe total, compared to our 1820 SNe. Within DES, we report 1623 likely ($P_{\rm SN Ia} >0.5$) SNe Ia, compared to 1635 SNe Ia in \citetalias{DES5YR}. At low-redshift, we have 197 SNe compared to 194 in \citetalias{DES5YR}. We find 1718 overlapping SNe between \citetalias{DES5YR} and DES-Dovekie, with approximately 100 different SNe between the two analyses. We find that supernovae passing fits with Dovekie not previously included in \citetalias{DES5YR} appear to be consistent with the overall $x_1/c$ populations; they are not bluer, redder, or from any particular stretch distribution. Table \ref{tab:distsigmas} shows the likelihood, presented in $\sigma$ discrepancy of a Kolmogorov-Smirnov (KS) test, that selected parameters are drawn from different distributions between \citetalias{DES5YR} and this work. 

\begin{table}
    \centering
    \begin{tabular}{l|ccc}
        \textbf{Parameter} & \multicolumn{3}{c}{\textbf{Difference Between DES-Dovekie and DES-SN5YR}}\\ 
         & $\sigma$ (DES) & $\sigma$ (Foundation) & $\sigma$(Low-z) \\
        \hline
         $z$ & $0.0\sigma$ & $0.0\sigma$ & $0.0\sigma$ \\
         $c$ & $2.8\sigma$ & $0.0\sigma$ & $0.2\sigma$ \\
         $x_1$ & $1.2\sigma$ & $0.2\sigma$ & $0.0\sigma$ \\
         $m_B$ & $0.3\sigma$ & $0.0\sigma$ & $0.0\sigma$ \\
    \end{tabular}
    \caption{KS test for consistency between DES-Dovekie and \citetalias{DES5YR}, for SALT3 parameter distributions
    and for each subsample. }
    \label{tab:distsigmas}
\end{table}

As expected, the distributions least sensitive to calibration ($x_1$, $z$) show no significant divergence at the population level. The only distribution that is significantly modified is the colour distribution among DES SNe; this is principally a shift in the mean of the distribution from $\bar{c}= 0.011$ to $\bar{c}=-0.003$. To compare the shapes of the colour distributions, we remove the $\bar{c}$ shift and redo the KS test; this reduces the  significance of the difference to $1.6\sigma$, as evaluated by bootstrap resampling. This colour shift is not surprising; colour is highly sensitive to calibration, and the mean of the colour distribution is determined by the demographics of the SALT training sample.  The calibration of the low-$z$ surveys is mixed, resulting in no coherent offset. The overall changes in distance are shown in Figure \ref{fig:HUBBLE}, but detailed further in Section \ref{sec:Results}. 

\subsection{Internal Consistency}

BBC relies on good agreement between the simulated bias corrections and observed data for an unbiased cosmology analysis. To accurately model intrinsic brightness in simulated bias corrections, \citetalias{DES5YR} used the best-fit parameters from {Dust2Dust} \citep{Popovic22}, which describes populations of stretch, colour, $M_{\star}$, $A_V$, and $R_V$. 

\begin{figure*}
    \centering
    \includegraphics[width=18cm]{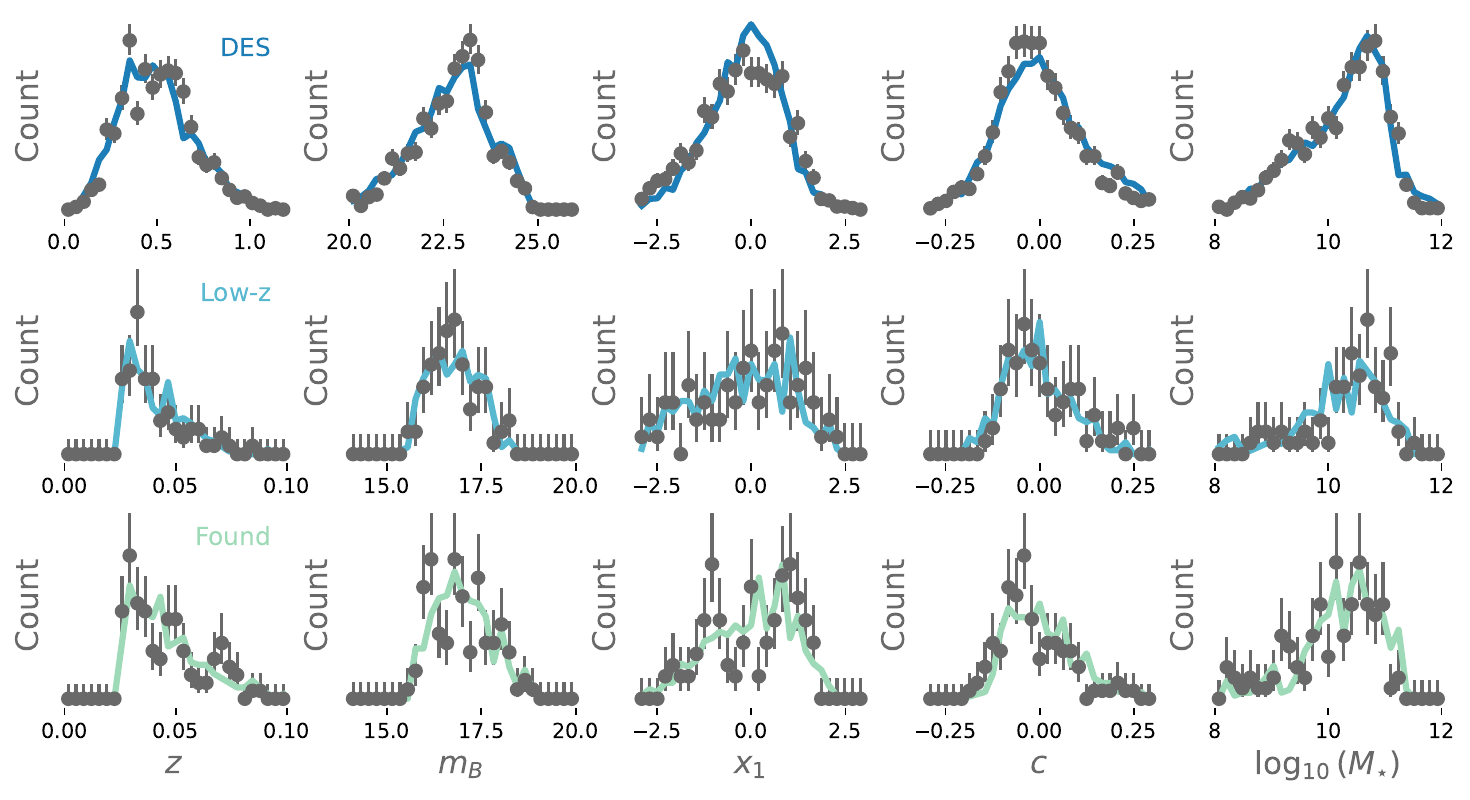}
    \caption{Comparison between the simulated and observed SN Ia parameters. The data are presented in grey, and the simulations are presented in coloured histogram. }
    \label{fig:DATASIM}
\end{figure*}

To check if the existing {Dust2Dust} model is valid in the DES-Dovekie analysis, we compare our data with our new SALT3.DOV simulations. Figure \ref{fig:DATASIM} shows the overlaid distributions of $z,m_B,x_1,c,M_{\star}$ for data and the new biasCor simulations; we see good agreement between data and simulations, and find no distribution that disagrees at more than $3\sigma$. Given the consistency of the data-sim comparisons, we have chosen to use the same {Dust2Dust} model that was used in \citetalias{DES5YR}.

The final consistency check is to test our ability to recover our simulated cosmology parameters from performing the full analysis on 25 statistically independent simulated samples. To avoid excessive CPU consumption using {Cosmosis} on 25 samples, we use a faster and simpler minimisation code\footnote{https://github.com/RickKessler/SNANA/blob/master/src/wfit.c} that replaces the full Planck likelihood with a parametrisation using the CMB R-shift parameter as described in \cite{Sanchez21}. Averaging results of the 25 samples for a Flat $w$CDM cosmology with a CMB prior, we recover our input cosmology, Flat $\Lambda$CDM with $\Omega_{\rm m} = 0.315$, to under $1\sigma$ for both parameters: $w_{\rm reco} = -0.9986   \pm 0.0045$ and $\Omega_{\rm m, reco} = 0.316 \pm 0.001$. When fitting for the $w_0w_a$CDM cosmology, we similarly recover our input cosmology using CMB priors: $w_{0, {\rm reco}} = -1.0027 \pm 0.0223$ and $w_{a, {\rm reco}} = -0.0325 \pm 0.1015$.

\begin{table}
    \centering
    \begin{tabular}{l|ccccc}
        Sample & $N_{\rm SNe}$ & $\alpha$ & $\beta$ & $\gamma$ & RMS  \\
        \hline
        Simulations & 1605 & 0.140 & 2.80 & 0.0 & 0.158 \\ 
        Data - \citetalias{DES5YR} & 1829 & 0.170(1) & 3.12(3) & 0.038(7) & 0.168  \\
        Data - DES-Dovekie & 1820 & 0.169(3) & 3.14(3) & 0.033(8) & 0.169 \\
        \hline
        \hspace{2mm} \textbf{DES} & 1679 & 0.17(1)  & 3.17(4) & 0.04(1) & 0.165 \\
        \hspace{2mm} \textbf{Foundation} & 118 & 0.15(1)  & 2.9(14)  & 0.015(23) &  0.110 \\
        \hspace{2mm} \textbf{Low-z} & 83 & 0.15(1)  & 2.88(14)  & -0.009(30) & 0.120 \\
    \end{tabular}
    \caption{Nuisance parameters from DES-Dovekie analysis, the original V24 analysis, and simulated data corresponding to DES-Dovekie}
    \label{tab:Nuisanceparams}
\end{table}

\subsection{Unblinding Criteria}

Through the course of this analysis, we blinded cosmological parameters estimated from real data until certain unblinding criteria were met. Pipeline validation, as in \citetalias{DES5YR}, was performed on realistic and detailed catalogue-level simulations, which were fit to test the recovery of simulated cosmology parameters. 

We follow the unblinding criteria laid out by \citetalias{DES5YR}:
\begin{itemize}

    \item \textbf{Recovery of original DES-SN5YR cosmology parameters:} To ensure that any measured changes in cosmology arise from the changes in Table \ref{tab:toplevel}, and not from changes in the pipeline since the publishing of \citetalias{DES5YR}, we repeat the original DES-SN5YR analysis using the same (current) {SNANA} code that is used in this reanalysis. For this test, a flag was added to the {SNANA} code to restore the \cite{Fitzpatrick99} colour law bug, and we restored the mistaken calibration weight.
    \item \textbf{Recovery of input simulated cosmology:} We produce 25 statistically-independent simulacra of the DES-SN5YR dataset assuming a Flat $\Lambda$CDM cosmology. Each of the 25 simulations is run through our analysis pipeline and fit with a CMB prior for $w$CDM and $w_0 w_a$CDM cosmology. 
    \item \textbf{Accuracy of simulations:} Following \citetalias{DES5YR}, test the accuracy of our simulations across our observables. Specifically, we test $z, x_1, c, m_B, M_{\star}, \mu - \mu_{\rm model}$, and the redshift and host-galaxy mass evolution of $x_1$ and $c$, and require a reduced $\chi^2$ between data and simulations to lay between 0.7 and 3.0. 
\end{itemize}

Additionally, we provide a brief summary of the consistency and improvement tests that \citetalias{Dovekie} performed during their analysis to ensure a good calibration solution:
\begin{itemize}
    \item \textbf{Recovery of Simulated Offsets:} \citetalias{Dovekie} simulated 100 catalogues of tertiary stars with a random ZP offset for each survey. They recovered their simulated offsets within $1\sigma$ across their 100 simulacra.
    \item \textbf{Updated DES-SN5YR Tertiary Stars:} \citetalias{Dovekie} increased the number of tertiary stars used for the cross calibration by a factor of 2.
    \item \textbf{Consistency between DA WD and Tertiary Stars Results:} The DES $g-r$ shift in Figure \ref{fig:deltazp} is consistent between the calibration results with DA WD modelling and with the conventional tertiary star approach. 
    \item \textbf{Hubble Residual Scatter:} \citetalias{Dovekie} found equal or improved HR scatter with their new calibration and SALT surface.
\end{itemize}

\section{Description of Systematic Uncertainties}\label{sec:systdesc}

Here we briefly summarise the sources of systematic uncertainty considered in DES-Dovekie. A more full description is available in \citetalias{DES5YR}; we follow their choices of systematic uncertainty. The covariance matrix for each systematic is calculated from Equation \ref{eq:covsyst}, and each $\cal C$ term is computed from a lightcurve fitting and BBC process. 

\subsection{Calibration}

To evaluate the calibration contributions to $\cal C$, we use the nominal SALT model with calibration from \citetalias{Dovekie}, to draw 9 systematic SALT surfaces randomly from the covariance matrix derived during the cross-calibration. Accordingly, we set $W_S = 1/\sqrt{9} = 0.33$, such that the quadrature sum of our SALT surface systematics is 1. 

An additional source of uncertainty, related to the CALSPEC flux calibration, is considered, with a shift of 5mmag/7000\r{A} \citep[as recommended in][]{Bohlin14} applied to the data.

\subsection{SN Ia Properties and Astrophysics}

The nuisance parameters $\alpha, \beta, \gamma$ in Equation \ref{eq:Tripp} are assumed to be constant in the nominal analysis. However, a redshift evolving nuisance parameter, such as $\alpha(z) = \alpha_0 + \alpha_1 \times z$, may capture un-modelled evolution of SN Ia parameters. As a systematic, we allow $\alpha$, $\beta$, and $\gamma$ to separately evolve with redshift with the same linear relation as above. We also perform a fit with  a fixed $\alpha =0.16, \beta=3.1$, to estimate the impact of the lower $\beta$ found in photometric data compared to spectroscopic data.

The location of the mass step is placed at $\log_{10}M_{\star} = 10$ in our nominal analysis. However, reflecting the uncertainty in both measurements of the stellar mass of host galaxies and the uncertainty of the nature of the mass step, we test the impact of moving the location of the mass step `split' to the median of the sample, at $\log_{10}M_{\star} = 10.3$.

During the Pantheon+ analysis, updates to the the intrinsic scatter $\sigma_{\rm int}$ definition were implemented, redefining it as 
\begin{equation}
         \sigma_{\mathrm{floor}}^2(z_i, c_i, M_{*,i}) = \sigma_{\mathrm{scat}}^2(z_i, c_i, M_{*,i}) + \sigma_{\rm gray}^2,
         \label{eq:sig_floor}
\end{equation}
which is a function of the $c,z,M_{\star}$ parameters, as opposed to the original BBC methodology which did not include $M_{\star}$-dependence.

{}

We revert to this constant $\sigma_{\rm int}$ model as a source of systematic uncertainty, to test the impact of this colour-dependent scatter term compared to the historical grey intrinsic scatter.

The largest source of systematic uncertainty in \citetalias{DES5YR} was the nature of SN Ia scatter. The \cite{BS20} model introduced the concept of SN Ia scatter being driven by differing dust distributions in the host galaxies of SNe Ia. Improved by \cite{Popovic22}, the systematic uncertainty test for \citetalias{DES5YR} was derived using three sets of dust parameters drawn from the chains of {Dust2Dust}. Additionally, a version of the \cite{BS20} model with adjusted parameters {(specifically the $\tau$ parameters do not match BS21)} was included, though the provenance of these changes has been lost. We include this systematic.

Finally, the nominal analysis uses the host galaxy stellar mass $M_{\star}$ as the tracer to explain correlations between the SNe Ia and their host environment. While the host galaxy stellar mass is the host property most robust to limited photometric information, it is not necessarily the most accurate tracer of SN Ia brightness. Following \cite{Wiseman22}, which proposes that the $u-r$ host galaxy colour may instead be the driving factor in these correlations, we incorporate an alternative model labelled as `W22', with a `colour step' analagous to the mass step. 

The W22 model utilises SNe rates and delay time distributions from \cite{Wiseman22} to drive galaxy evolution including stellar age, mass, and star formation rate, which is then correlated to the SNe Ia properties via the relationships via \cite{Rigault2020}.

\subsection{Milky Way Extinction}

Given the importance of the dust distributions of other galaxies, we consider the impact of our own Milky Way galaxy. The SALT3 fitting code includes Galactic extinction in the model fluxes; inaccurate extinction modelling can bias the SALT3-fitted colours.

We consider two systematics associated with the Milky Way corrections. The first is a global scaling of 5\% down {of the \cite{Schlafly11} values, following \cite{Brout22}}. The second systematic is a change from the \cite{Fitzpatrick99} reddening law to that presented from \cite{Cardelli89}. 

\subsection{Host and Survey Modeling}

Simulations of SNe Ia within the SNANA framework draw from a realistic catalogue of galaxies to model the observed SN-host correlations in the data. While the nominal catalogue is generated from DES co-added images from \cite{Wiseman20,Qu23}, we test an additional, shallower catalogue (`SVA Gold'), based on DES science verification data, used in other DES analyses: \cite{Smith20,Kessler19}.

Further, we test the systematics of our selection function. We model the efficiency of obtaining the spectroscopic redshift of a host galaxy as a function of host galaxy brightness, $\epsilon_z^{\rm spec}$ \citep{Vincenzi21}, which helps model the selection effects of the DES survey. The systematic test we apply is to shift this function to observe host galaxy $r-$band brightness that are $+0.2$ and $-0.2$ magnitudes fainter and brighter, respectively. Their weights are $W_S = \sqrt{1/2}$ each. 

\subsection{Contamination and Photometric Classifiers}

Compared to analyses with spectroscopically-confirmed SNe Ia, DES included non-Ia contamination within the sample, potentially biasing the distances via the use of non-standardisable SNe. The nominal analysis uses the SuperNNova classifier by \cite{Moller19}, which had been rigorously tested by \cite{Vincenzi21}. 

The systematic analysis of the core collapse contamination comes in three parts: simulated Ia+non Ia training set, modelling the CC prior in BBC with simulated sample, 
and the classifier method.

The nominal analysis used the simulations of non-Ia SNe developed by \cite{Vincenzi19}; this is replaced for systematic tests by the templates and simulations from \cite{Jones16}, hereafter J17, and additionally real-data observed in DES, hereafter "DES-CC".

To evaluate the uncertainty on classification method, SNN was replaced with two alternate classifiers: SCONE \citep{Qu21} and the Supernova Identification with Random Forest (SNIRF). These classifiers are trained on the same set of simulations that the nominal SuperNNova model was trained on. Additional training sets were also used for systematics.

Each of these approaches relies on a suite of simulations of non-Ia SNe to inform the BEAMS method about populations; a final systematic test for BBC is to replace the simulated non-Ia prior with a redshift-dependent polynomial fit as described in  \cite{Hlozek12}. 

\subsection{Redshift}

A coherent redshift shift of $4\times 10^{-5}$ is applied to the data, following \cite{Calcino17}, to test for the impact of a local void or other such redshift errors.

To correct for peculiar velocities, we use corrections from 2M++; our systematic uncertainties come from approaches implemented in \cite{Peterson21}. The first is to maintain the 2M++ corrections, but integrate over the line-of-sight between the distance and the redshift. Secondly, we switch to the peculiar velocity map from 2MRS \citep{Lilow21}. Both systematics are weighted such that they sum in quadrature to 1. 

\section{Results}\label{sec:Results}

\subsection{Reproducing DES-SN5YR}

The results of \citetalias{DES5YR} were unblinded in 2023, and in the intervening years, {SNANA} has introduced updates to the lightcurve fitting, simulating, and cosmology-fitting programs that were originally used in \citetalias{DES5YR}. To check if the changes in cosmology that we measure are solely due to changes in Table \ref{tab:toplevel}, and to ensure consistency in our pipeline across time, we attempt to replicate the published results of \citetalias{DES5YR}. This attempt to recreate DES-SN5YR `as-published', dubbed `Lyrebird', will be used to track and identify major changes to {SNANA} that have shifted the final cosmology results. This Lyrebird test uncovered two additional changes: a code fix to the \cite{Fitzpatrick99} colour law (Appendix \ref{sec:F99}), and an input mistake in the systematic weight for calibration. The Lyrebird test includes the F99 code bug and the incorrect calibration weight. 

Because of the potential code and pipeline changes, and the use of up-to-date data and likelihoods for the CMB and BAO, we do \textit{not} expect Lyrebird to recover the exact same cosmological parameters as \citet{DES5YRKP}. However, we have checked that our Lyrebird results are compatible with the results published in the DES Key Paper and DESI DR2 \citep{DESIDR2}. A summary of our recovered Lyrebird cosmology is given in Table \ref{tab:Lyrebirdcosmo}; for this table, we take the published \citetalias{DES5YR} distances and covariance and use a quick-fitting solution, replacing the full CMB likelihood with distance priors derived from the results of \cite{Lemos2023}.  We find agreement in all cases within $\ll 1\sigma$. 

\begin{table}
    \centering
    \begin{tabular}{c|cc}
        Parameter & Original \citetalias{DES5YR} Distances & Lyrebird  \\
        \hline
        $\Omega_{\rm m}$ (Flat $\Lambda$CDM) & $+0.352 \pm 0.017$ & $+0.357 \pm 0.017$ \\ 
        \hline
        $w$ (Flat $w$CDM) & $-0.934 \pm 0.027$ & $-0.927 \pm 0.027$ \\
        $\Omega_{\rm m}$ (Flat $w$CDM) & $+0.330 \pm 0.009$ & $+0.331 \pm 0.008$ \\ 
        \hline
        $w_0$ (Flat $w_0w_a$CDM) & $-0.770 \pm 0.060$ & $-0.760 \pm 0.060$ \\ 
        $w_a$ (Flat $w_0w_a$CDM) & $-0.740 \pm 0.220$ & $-0.750 \pm 0.230$ \\ 
        $\Omega_{\rm m}$ (Flat $w_0w_a$CDM) & $+0.318 \pm 0.006$ & $+0.319 \pm 0.006$ \\ 
        
    \end{tabular}
    \caption{Comparison of cosmological parameters, between the published \citet{DES5YRKP} distances and the recreation of those distances, nicknamed Lyrebird in this work. We pick data combinations that are most informative for comparison : in Flat $\Lambda$CDM, SN Ia only data is used, Flat $w$CDM  uses SN+CMB, and Flat $w_0 w_a$CDM uses SN+CMB+DESI BAO. The values for the original distances are not expected to exactly match those in \citet{DES5YRKP} as the BAO and CMB likelihoods have been updated, and fitting uses the {Nautilus} sampler. The minor changes in sample composition do not affect the cosmology solution.}
    \label{tab:Lyrebirdcosmo}
\end{table}

\subsection{The DES-Dovekie Hubble Diagram}

Figure \ref{fig:HUBBLE} presents the DES-Dovekie Hubble diagram. We find 1623 DES likely SNe Ia, and 197 SNe Ia at low redshift. Our low-redshift sample has 3 additional events compared to \citetalias{DES5YR}, and our DES sample has 9 fewer events. Table \ref{tab:cuts} provides a breakdown on the quality and survey cuts we perform.

\begin{table}
    \centering
    \begin{tabular}{l|ccc}
         & \multicolumn{3}{c}{\textbf{DES SN}}   \\
         Requirement & \textbf{Low-z} & DES & Total  \\ 
         \hline
          SALT3 fit & 376 & 3590 & 3991  \\ 
          $|x_1 <3| \& |c<0.3|$ & 313 & 2818 &  3196 \\ 
          $\sigma_{x_1} <1.15$, $\sigma_{t_{\rm peak}} <2$ & 304 & 2259 &  2628 \\ 
          Valid Host $z$ & 291 & 1710 &  2066 \\
          Chauvenet's Criterion & 207 & 1682 &  1943 \\
          Valid Bias Corrections & 201 & 1680 & 1881 \\ 
          Common SNID & 197 & 1623 &  1820  \\ 
          \hline
          Total & 197 & 1623 &  1820 \\ 
    \end{tabular}
    \caption{Number of SN Ia remaining after quality and survey cuts.}
    \label{tab:cuts}
\end{table}

Table \ref{tab:Nuisanceparams} provides the final nuisance parameters $\alpha$, $\beta$, $\gamma$, and the Hubble residual RMS, for the DES-Dovekie data and 25 simulated data sets, alongside a summary of the equivalent values for \citetalias{DES5YR}. While we have a slightly smaller sample, the nuisance parameters remain unchanged between \citetalias{DES5YR} and DES-Dovekie. Nonetheless, analysis of DES-Dovekie simulations maintain the same curious $\beta = 2.8$ as \citetalias{DES5YR}.

\begin{table*}
\centering
\begin{tabular}{lclc}
\Large{Baseline} & \Large{Weight} & \Large{Systematic} & \Large{Label} \\ 
\hline
\multicolumn{4}{l}{\textbf{Calibration and Light-curve Modeling }}\\
\hspace{2mm} \color{MidnightBlue}{SALT3 surfaces $\&$ ZP} & 1/10 & 10 covariance realizations  & \lq SALT3+Calibration\rq \\
\hspace{2mm} HST Calspec 2020 Update & 1 & 5 mmag/7000\AA\ & \lq HST Calspec\rq  \vspace{2mm}\\
\multicolumn{4}{l}{\textbf{SN Ia properties and astrophysics }}\\
\hspace{2mm} \color{MidnightBlue}{Dust-based model \cite{Popovic22} (\lq P23($M_{\star}$)\rq)} & 1/3 & 3 realizations from MCMC dust model fitting code & \lq P23 dust pop 1/2/3\rq \\
\hspace{2mm} & 1 & Original BS20 dust parameters & \lq BS21\rq \\
\hspace{2mm} & 1 & Splitting on $u-r$ & \lq P23($u-r$)\rq \\
\hspace{2mm} \color{MidnightBlue}{Empirical modeling of $x_1$-M$_{\star}$ correlations} & 1 & Modeling SN Ia age following \citet{Wiseman22} & \lq Model SN Ia age\rq\\
\hspace{2mm} Fixed $\alpha$/$\beta$ & 1 & $\alpha = 0.16, \beta = 3.1$ & \lq Fixed $\alpha/\beta$\rq \\
\hspace{2mm} No $\alpha$ evolution & 1 & $\alpha(z) = \alpha_{0} + \alpha_{1}\times z$ & \lq $\alpha$ Evolution\rq \\
\hspace{2mm} No $\beta$ evolution & 1 & $\beta(z) = \beta_{0} + \beta_{1}\times z$ & \lq $\beta$ Evolution\rq \\
\hspace{2mm} No $\gamma$ evolution & 1 & $\gamma(z) = \gamma_{0} + \gamma_{1}\times z$ & \lq $\gamma$ Evolution\rq \\
\hspace{2mm} Mass step location at $10^{10} M_{\odot}$& 1 & $10^{10.3} M_{\odot}$ & \lq Mass Location\rq \\
\hspace{2mm} $\sigma_{\rm int}$ modeling with scaling+additive scatter terms (Eq. \ref{eq:sig_floor}) & 1 & Scaling term only & \lq $\sigma_{\rm int}$ modeling\rq \vspace{2mm}\\
\multicolumn{4}{l}{\textbf{Milky Way extinction }}\\
\hspace{2mm} MW scaling \citet{Schlafly11} & 1 & 5\% scaling & \lq MW scaling\rq \\
\hspace{2mm} MW color law $R_V$=3.1 and F99 & 1/3 & $R_V$=3.0 and CCM &\lq MW color law\rq\vspace{2mm}\\
\multicolumn{4}{l}{\textbf{Host and survey modeling }}\\
\hspace{2mm} \color{MidnightBlue}{SN Ia host catalog by \citet{Qu23}} & 1 & SN Ia host catalog using SVA Gold galaxy catalog &\lq DES SV catalog\rq \\
\hspace{2mm} \color{MidnightBlue}{Efficiency $\epsilon_{z}^{\mathrm{spec}}$ presented by \citetalias{Vincenzi20}}  & 1 & Shift of $\pm$0.2 mag in the efficiency curves &\lq Shift in host spec eff\rq \vspace{2mm}\\
\multicolumn{4}{l}{\textbf{Contamination and photometric classifiers }}\\
\hspace{2mm} Classification using SuperNNova & 1 &  SCONE, SNIRF \\
\hspace{2mm} {Classifier training sample simulated using \citetalias{Vincenzi19} templates} & 1 & \citetalias{Jones17} templates, DES CC templates (\lq {SuperNNova} training\rq )\\
\hspace{2mm} Core-collapse SN prior using \citetalias{Vincenzi19} simulation & 1 & Polynomial fit as in \citet{Hlozek12} &\lq CC SN prior\rq \vspace{2mm}\\
\multicolumn{4}{l}{\textbf{Redshift }}\\
\hspace{2mm} Peculiar velocities using 2M$++$ & 1 & 2M$++$(Line-of-sight integration) or 2MRS &\lq Pec Velocities\rq \\
\hspace{2mm} No redshift shift & 1/6 &  $\Delta z = 4 \times 10^{-5}$ &\lq Redshift shift\rq \\

\end{tabular}
\caption{An overview of the systematics in DES-SN5YR and this reanalysis. Those systematics that are directly impacted by the recalibration in \citetalias{Dovekie} are shown in \color{MidnightBlue}{blue}. }
\label{tab:syst_description}
\end{table*}

\begin{center}
\input{SystUncertainty}
\end{center}

\section{Systematic Uncertainties}\label{sec:SYST}

We assess the systematic uncertainties here, and compare to those in \citetalias{DES5YR}, with the systematic uncertainties impacted by the \citetalias{Dovekie} calibration highlighted in \textcolor{MidnightBlue}{Midnight Blue}. As per \citetalias{DES5YR}, we analyse systematics in the context of the Flat $w$CDM model, using the fast cosmology fitter {wfit}, returning to more sophisticated models and fitters for the cosmology solution.

Table \ref{tab:syst_size} gives the values of the individual systematic uncertainties, for SNe alone without any priors. As in \citetalias{DES5YR}, we remind the reader that the individual systematic uncertainties \textit{do not} sum to the total systematic uncertainty. We evaluate each individual contribution with a \say{leave-one-out} approach, comparing the uncertainty with and without a given systematic. With the all sources of systematic uncertainty considered, internal correlations between systematics partially cancel out during the fitting process. This is not the case during the leave-one-out approach, and therefore the systematic uncertainties do not sum to the total systematic uncertainty. In a similar vein, the $\delta w ~ (w_{\rm stat+syst} - w_{\rm stat})$  values in Table \ref{tab:syst_size}, which are generated with an \say{add-one-in} approach, do not match the total change in $w$. Remembering the internal correlations within the cosmology fitting process, it is not surprising that several of the systematic uncertainties have shifted with regards to \citetalias{DES5YR}.

\begin{table*}
\centering
\caption{Nuisance parameters and quality of fit changes.}
\label{tab:systematic_uncertainties}
\begin{tabular}{l|ccc|cc|cc}
\toprule
\multirow{2}{*}{Systematic} & \multicolumn{3}{c|}{BBC Parameters} & \multicolumn{2}{c|}{Fit Quality} & \multicolumn{2}{c}{Cosmology} \\
\cmidrule(lr){2-4} \cmidrule(lr){5-6} \cmidrule(lr){7-8}
 & $\alpha$ & $\beta$ & $\gamma$ & $\sigma_{\rm int}$ & RMS & $\delta \chi^2$ & $\Delta w$   \\
\midrule
None & 0.169(3) & 3.14(4) & 0.033(8) & 0.035 & 0.170 & 0.0 & 0.000 \\
\midrule
BS21 & 0.169(4) & 3.21(4) & 0.021(9) & 0.053 & 0.170 & -1.5 & -0.003 \\
DUST1 & 0.168(3) & 3.21(4) & 0.041(8) & 0.050 & 0.170 & 1.0 & +0.000 \\
DUST2 & 0.167(3) & 3.08(4) & 0.026(9) & 0.030 & 0.171 & 7.0 & +0.003 \\
DUST3 & 0.167(4) & 3.09(4) & 0.037(9) & 0.046 & 0.171 & 23.2 & -0.001 \\
Model SN Ia Age (W22) & 0.168(4) & 3.12(4) & 0.038(8) & 0.045 & 0.170 & 1.4 & -0.002 \\
New $\alpha$ / $\beta$ guess & 0.179(3) & 3.37(4) & 0.034(8) & 0.039 & 0.170 & -0.5 & 0.000 \\
\addlinespace[0.5em]
$\alpha$ Evolution & 0.164(6) & 3.15(4) & 0.033(8) & 0.030 & 0.170 & 1.1 & +0.001 \\
$\beta$ Evolution & 0.169(3) & 3.08(8) & 0.033(8) & 0.035 & 0.170 & 2.1 & -0.006 \\
$\gamma$ Evolution & 0.169(3) & 3.15(4) & 0.04(2) & 0.035 & 0.170 & -1.6 & +0.000 \\
\addlinespace[0.5em]
$\sigma_{\rm int}$ model & 0.168(4) & 3.13(4) & 0.031(9) & 0.113 & 0.170 & 3.0 & 0.000 \\
DES SV Catalogue & 0.16813(1) & 3.1248(3) & 0.035(8) & 0.036 & 0.170 & 3.2 & +0.001 \\
INTRSC COLOUR & 0.164(4) & 3.15(4) & 0.029(8) & 0.050 & 0.172 & 38.4 & +0.005 \\
MW Scaling & 0.169(4) & 3.14(4) & 0.035(8) & 0.030 & 0.170 & 2.1 & -0.003 \\
HST CALSPEC & 0.168(4) & 3.15(4) & 0.034(8) & 0.036 & 0.170 & -1.1 & -0.000 \\
MW Colour Law  & 0.170(4) & 3.14(4) & 0.035(8) & 0.030 & 0.170 & -0.7 & -0.003 \\
\bottomrule
\end{tabular}
\begin{tablenotes}
\small
\item Note: Errors on $\alpha$, $\beta$, and $\gamma$ are shown in parentheses.
\item $\Delta w$ is the $w$-shift relative to the baseline without systematic uncertainties using a \say{add-one-in} approach.
\item $\delta \chi^2$ shows the change in $\chi^2$ relative to baseline.
\item {$\sigma_{\rm int}$ is calculated as in Equation \ref{eq:sig_floor} such that the reduced $\chi^2 = 1$.}
\end{tablenotes}
\end{table*}

\begin{figure}
    \centering
    \includegraphics[width=8cm]{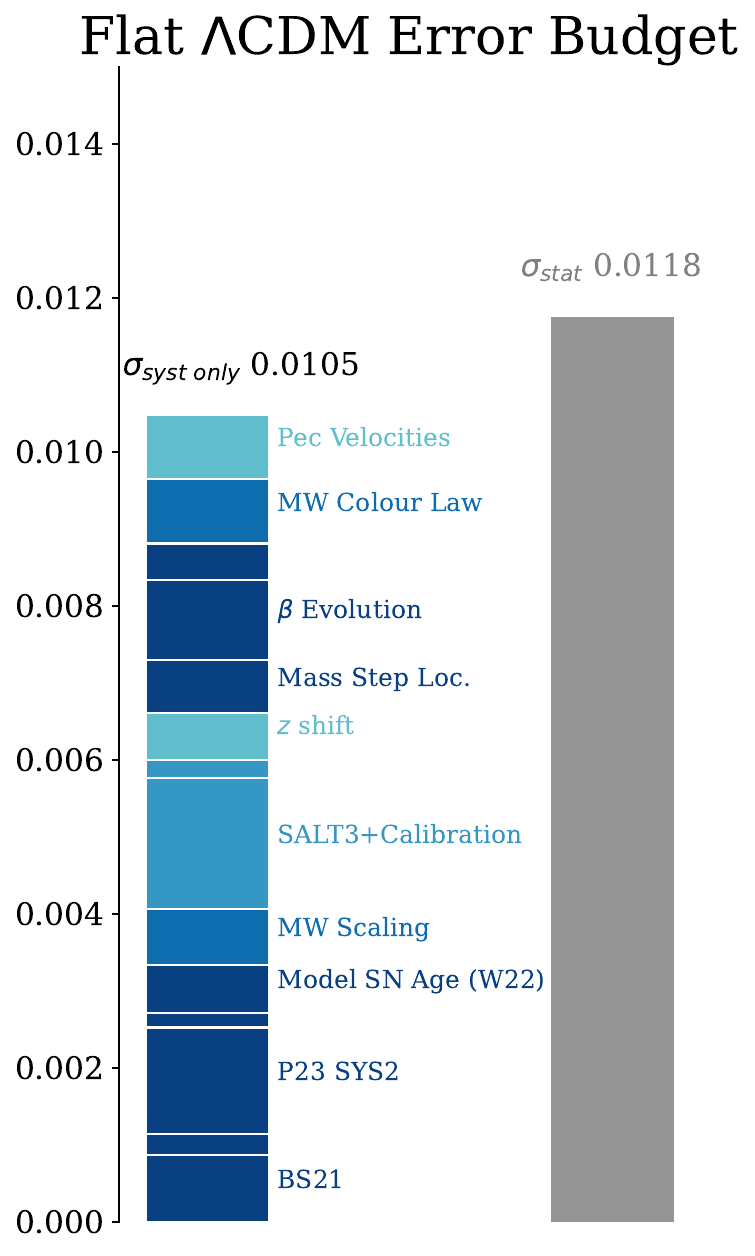}
    \caption{Systematic and statistical error budget on $\Omega_{\rm m}$ for a Flat $\Lambda$CDM cosmology with SNIa only. Systematics are colour coded as in Figure \ref{fig:ERR_BUDGET}, and only significant ($\sigma_{\rm sys} > 0.005$) systematics are labelled. The statistical error is presented in grey. }
    \label{fig:OM_ERR_BUDGET}
\end{figure}

\begin{figure}
    \centering
    \includegraphics[width=10cm]{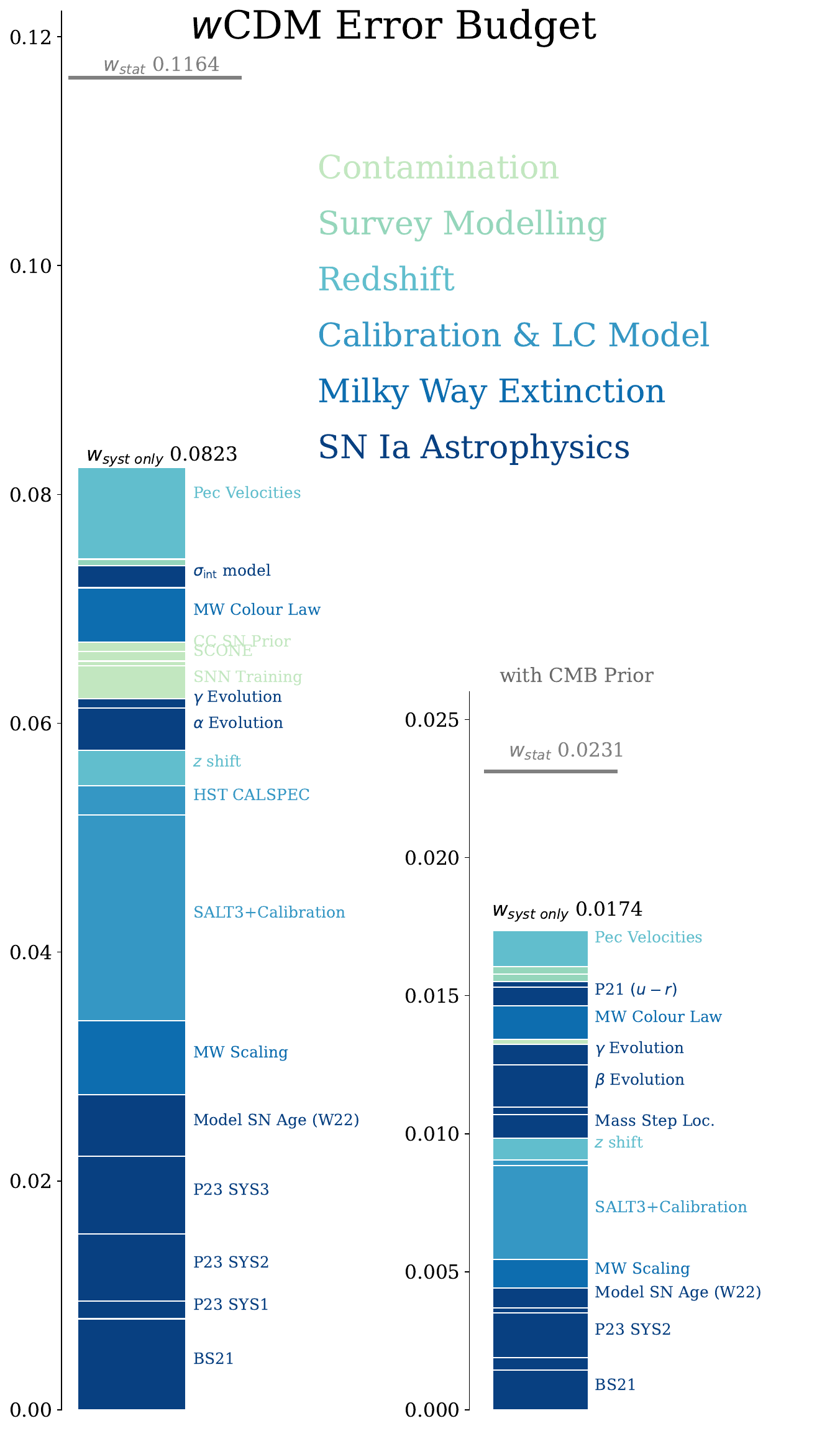}
    \caption{Systematic and statistical uncertainty budget on $w$, both with (right) and without (left) a CMB prior. {Each box is scaled by the total systematic uncertainty.} Those systematic uncertainties that are not labelled with text represent a negligible (<0.005) systematic contribution.}
    \label{fig:ERR_BUDGET}
\end{figure}

Firstly, we show the error budget for $\Omega_{\rm m}$ in a Flat $\Lambda$CDM universe for SNe Ia alone in Figure \ref{fig:OM_ERR_BUDGET}. We present the equivalent error budget but for Flat $w$CDM in Figure \ref{fig:ERR_BUDGET} with and without a CMB prior, which we shall focus on for this section. 

\subsection{\textcolor{MidnightBlue}{Calibration and Light-curve Modeling}}

We find that the combined systematic uncertainty that describes calibration and SALT training, $\sigma_w(\rm phot)$ has increased from \citetalias{DES5YR}, going from 0.057 to 0.075 in our analysis. However, the comparable $\sigma_w(\rm phot)$ values hides noticeable changes between this work and \citetalias{DES5YR}. The relative weights from \citetalias{DES5YR} have been changed to no longer underestimate the photometric calibration uncertainties (as in Section \ref{sec:Differences}). We find a SALT3+Calibration uncertainty in $w$ of only 0.066, compared to 0.052 in \citetalias{DES5YR}. Taking into account the increase of $\sim20\%$ from the change in relative weights, we see the Dovekie photometric uncertainty is consistent with \citetalias{DES5YR}.

\begin{figure}
    \centering
    \includegraphics[width=9cm]{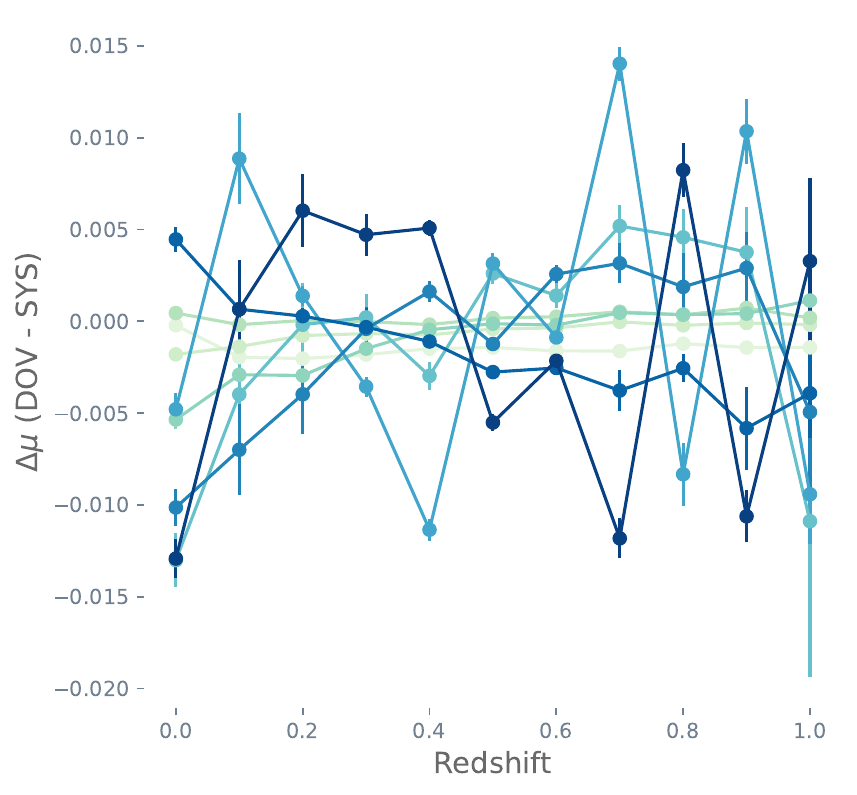}
    \caption{The median binned differences between the nominal distances and the 9 calibration systematics (colour coded for each systematic surface) for DES-Dovekie. Comparable to Figure 10 in \citetalias{Dovekie}, though we observe no obvious redshift-dependence.}
    \label{fig:CALIB}
\end{figure}

Figure \ref{fig:CALIB} shows the impact on the bias-corrected $\mu$ for the 9 sets of distance moduli used to assess the systematic uncertainty. We see no obvious trend with redshift. 

\subsection{\textcolor{MidnightBlue}{SN Ia Properties and Astrophysics}}

SN Ia properties and astrophysics remains the largest source of systematic uncertainty to-date, approximately $\times 2$ the size of Calibration and LC modelling. We find a slight decrease of 0.009 in $\sigma_w(\rm Astro)$ over \citetalias{DES5YR}, though this is not driven by any particular one systematic. 

\subsubsection{Dust Systematics}

The dust systematics -- P23 dust pop 1-3, P23$(u-r)$, BS21, and W22 -- represent the largest grouping of systematic uncertainties for astrophysics in SNe Ia at $\sigma_{w}(\rm dust) = 0.101$. Table \ref{tab:systematic_uncertainties} shows the $\delta \chi^2$ for these systematics; none of them are particularly favoured over the nominal P23 model. Noticeably, the largest $\delta w = 0.029$ is from the P23$(u-r)$, which has no contribution to the total systematic uncertainty -- likely indicating that this particular systematic does not well-match the data.

\subsubsection{Nuisance Parameter Systematics}

The remaining properties/astrophysics systematics relate to SN Ia nuisance parameters $\alpha,\beta,\gamma,\sigma_{\rm int}$ and the changes to their redshift evolution and initial properties. Interestingly, there is a slight preference to a $\gamma(z)$ at $-1.6\chi^2$, and a $-0.5\chi^2$ change when changing our initial estimations of $\alpha$ and $\beta$. Overall, these nuisance parameters contribute to our error budget but are subdominant to the suite of dust systematics.

\subsection{Milky Way Extinction Systematics}

We find a marginal increase to the Milky Way Extinction systematics; given the consistency between Milky Way Extinction systematics between \citetalias{Dovekie} and \citetalias{fragilistic}, this is not surprising. 

\subsection{Host and Survey Modelling Systematics}

\begin{figure}
    \centering
    \includegraphics[width=8cm]{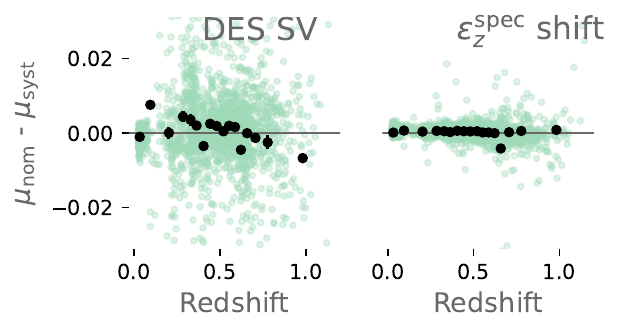}
    \caption{Effects of different Survey modelling systematics on the inferred SN Ia distances. Left: Choosing a shallower galaxy catalogue (DES `SVA Gold’). Right: Varying the efficiency of obtaining a spectroscopic redshift. Green points are individual realisations, black points are binned means.}
    \label{fig:HOSTSYST}
\end{figure}

Figure \ref{fig:HOSTSYST} shows the two `Survey modeling' systematics, changing the host spectroscopic efficiency `Shift $\epsilon_{z}^{\rm spec}$' and swapping the host library for the SVA Gold catalogue. We find the systematic uncertainty from both tests is consistent with 0. Figure \ref{fig:HOSTSYST} shows the $\Delta \mu$ for these host systematics.

\subsection{Contamination and Photometric Classifier Systematics}

Following Milky Way Extinction and Survey Modelling, we find small reductions to the systematics associated with contamination. The total systematic uncertainty is marginally smaller than \citetalias{DES5YR}, and there appears to be some migration between the sources of uncertainty - our uncertainty due to SuperNNova alternative training is now 0.010, compared to 0.006 in \citetalias{DES5YR}.

\subsection{Redshift Systematics}

Finally, we consider our redshift systematics, which have slightly increased from \citetalias{DES5YR}; this change is driven by an increased uncertainty from the peculiar velocities.

\section{Cosmology}\label{sec:Cosmo}

Here we present our constraints on cosmological parameters for DES-Dovekie (our nominal data set with updated calibration and colour law) for four models in particular: Flat $\Lambda$CDM, 
$\Lambda$CDM, Flat $w$CDM, and Flat $w_0w_a$CDM, with four different sets of probes: SN-only, SN+CMB, SN+BAO and SN+CMB+BAO. We reviewed our external probes in Section \ref{sec:Methodology}, but as a reminder CMB refers to the combination of Planck \citep{Planck20}, ACT \citep{ActDr6like} and SPT \citep{Camphuis2025} temperature and polarisation spectrum (TTTEEE) and lensing reconstruction data, and BAO refers to DESI DR2 measurements as described in Table IV of \citet{DESIDR2}. We summarise our results in Table \ref{tab:cosmology}. The values we quote are medians throughout, with the error bars representing the 16\% and 84\% percentiles. 

In Appendix \ref{sec:F99Cosmo}, we present the changes to \citet{DES5YRKP} arising \textit{solely} from the changes to the F99 colour law. Comparisons of DES-Dovekie to DES-SN5YR are shown in Appendix \ref{sec:AdditionalCosmo}.

\subsection{{Data consistency}}

It is well-known that combining data that is in tension (according to some suitable metric) within a given model results in artificially tight parameter constraints in the standard analysis. We check that DES-Dovekie is compatible with CMB, BAO and the BAO+CMB combination using the Suspiciousness statistic as described in \citet{Handley2019}. Suspiciousness is defined as 
\begin{equation}
    \log S = \log R - \log I \;,
\end{equation}
where the $R$ statistic
\begin{equation}
    \log R = \log \mathcal{Z}_{AB} - \log \mathcal{Z}_{B} - \log \mathcal{Z}_{B} 
\end{equation}
is defined from the relative evidences of the combination of datasets $A$ and $B$, that is the ratio of the probability of $B$ knowing $A$ compared to the probability of $B$ (in a given model). The evidence $\mathcal{Z}$ was defined in equation \ref{eq:evidence}. The $R$ statistic is prior-dependent so the term
\begin{equation}
\log I = \mathcal{D}_{A} +\mathcal{D}_{B} -\mathcal{D}_{AB} 
\end{equation}
is subtracted to remove this. $\mathcal{D}_{A}$ is the Kullbeck-Leibler divergence of dataset $A$
\begin{equation}
\mathcal{D}_{A} = \Bigl\langle \log \frac{\mathcal{P}_{A}}{\pi} \Bigr\rangle_{\mathcal{P}_{A}} \;,
\end{equation}
which quantifies the gain in information from the prior parameter distribution $\pi$ to the posterior $\mathcal{P}$. We choose $S$ as it is defined via integrals in information-theoretic terms, and is therefore re-parameterisation invariant (and so robust in the case of non-Gaussian distributions, which an important feature here). It also works with the entire parameter space rather than a subset, so tensions cannot be concealed or exaggerated by projection effects.

We use the criteria described in \citet{Handley2019} to evaluate a data combination : $\log S > -2.5$ is compatible, $-5 < \log S < -2.5$  is moderate tension, and $\log S < -5$ is strong tension. The $S$ statistic is $\chi^2$-distributed so we also convert $S$ to an equivalent $n\sigma$.

\subsection{Flat $\Lambda$CDM}\label{sec:Cosmo:subsec:LCDM}

\begin{figure}
    \centering
    \includegraphics[width=8cm]{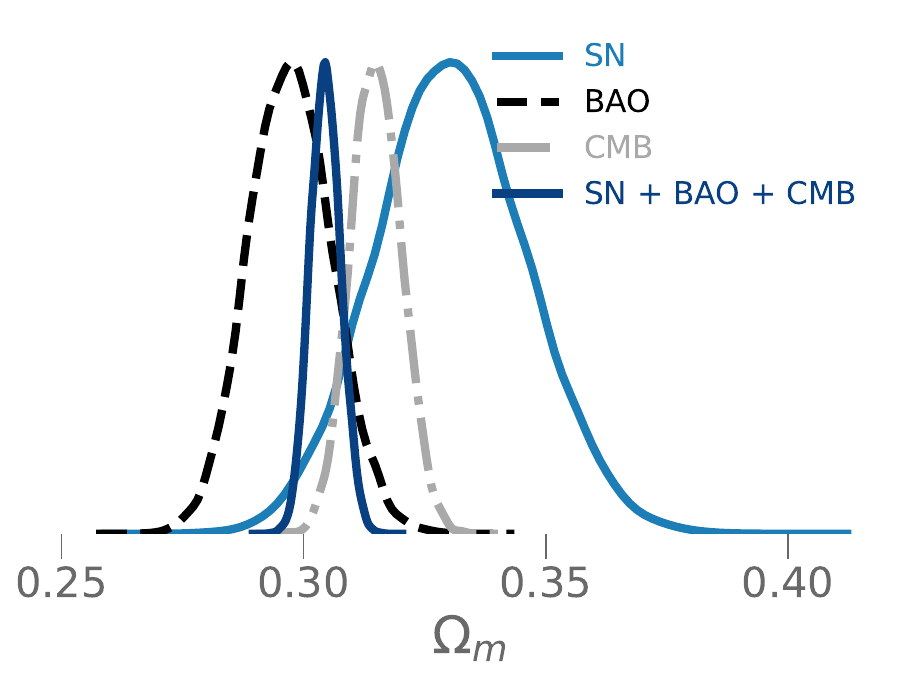}
    \caption{Constraints on the matter density for Flat $\Lambda$CDM for DES-Dovekie (blue), Planck CMB (grey), DESI DR2 BAO (black), and the combination SN+CMB+BAO (dark blue).}
    \label{fig:OM-FLCDM}
\end{figure}

In Figure \ref{fig:OM-FLCDM}, we show the $\Omega_{\rm m}$ posteriors for DES-Dovekie, CMB and BAO, with data combinations presented in Table \ref{tab:cosmology}. For DES-Dovekie, we find an $$\Omega_{\rm m} = 0.330 \pm 0.015~(\rm SN~only),$$ which is consistent with CMB results. Adding CMB data results in $$\Omega_{\rm m} = 0.317 \pm 0.005~ (\rm SN+CMB)$$ which is $1.9\sigma$ from DESI DR2 which has $\Omega_{\rm m} = 0.297 \pm 0.009$. 

{For the combination of DES-Dovekie with BAO+CMB, $\log S = -1.05$ which is a p-value equivalent to $1.7\sigma$.} Hence we judge it reasonable to combine all three datasets which results in $$\Omega_{\rm m} = 0.3045 \pm 0.0032 ~ (\rm SN+CMB+BAO).$$ 

With the addition of CMB data, we also find the Hubble constant as $H_0 = 68.14 \pm 0.23$ km sec$^{-1}$ Mpc$^{-1}$, and discuss this further in Section \ref{sec:hubble}.

\subsection{Open $\Lambda$CDM}\label{sec:Cosmo:subsec:oLCDM}
Fitting DES-Dovekie to a universe with non-zero spatial curvature, we find $$\Omega_{\rm k} = 0.14 \pm 0.15 \;,$$ where for this constraint we widen our priors to encompass the full posterior. For the SN+BAO combination we have $\Omega_{\rm k} = 0.054^{+0.033}_{-0.036}$, and SN+CMB gives $\Omega_{\rm k} = -0.0063 \pm 0.0037$. The full combination of SN+BAO+CMB results in $$\Omega_{\rm k} = 0.0026 \pm 0.0011~(\rm SN+BAO+CMB).$$, and $\log S=-0.97$ for DES-Dovekie compared to BAO+CMB.  

\begin{figure}
    \centering
    \includegraphics[width=9cm]{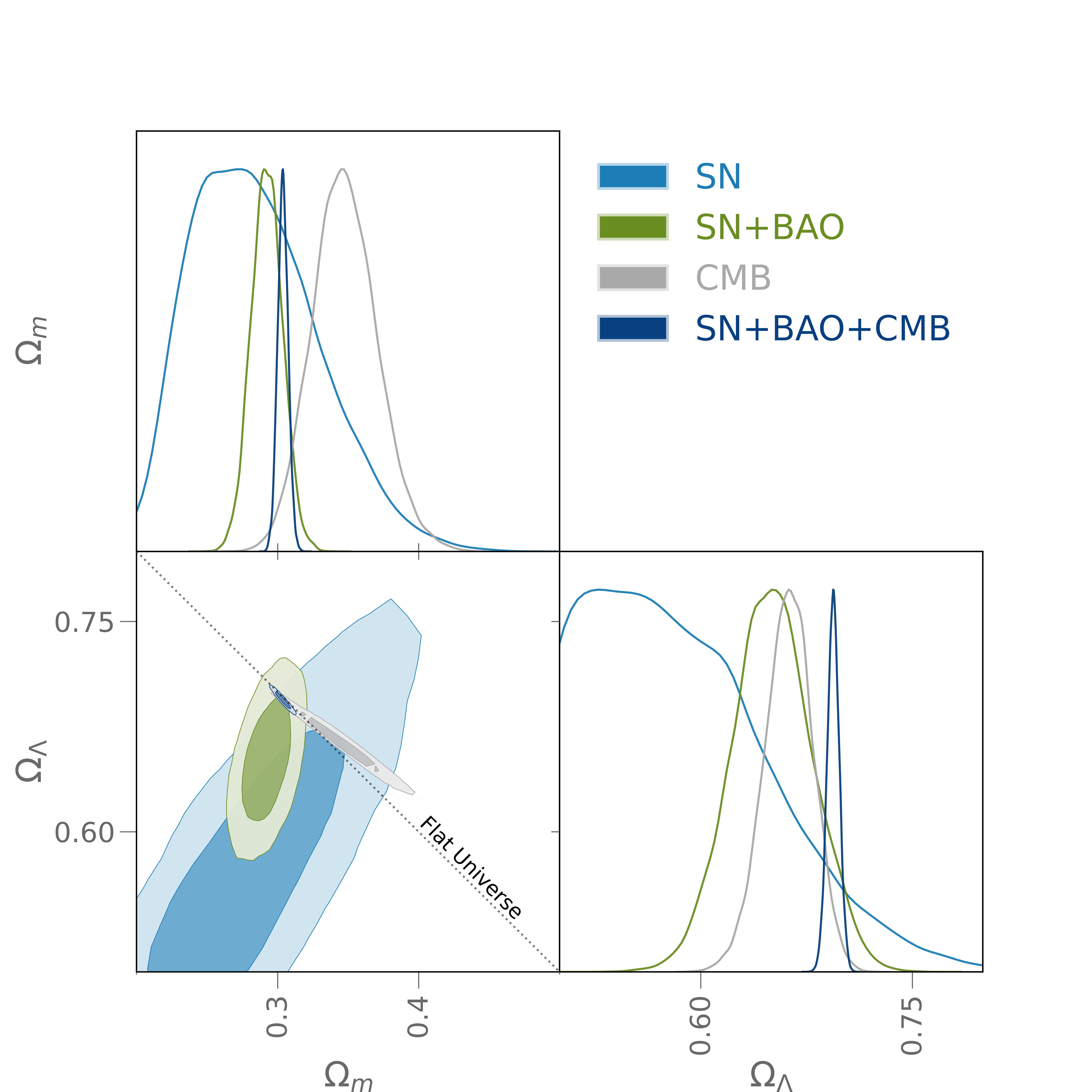}
    \caption{The $\Omega_{\rm m},\Omega_{\Lambda}$ contours for open $\Lambda$CDM. DES-Dovekie contours are in blue, complemented by the CMB (grey) and SN+BAO (olive) constraints. The full combination of SN+BAO+CMB is in dark blue. As a visual guide, we show the flat universe $\Omega_{\rm m}+\Omega_{\Lambda}=1$ as a dotted line. Although the SN contours appear truncated in this figure (to aid seeing the SN+BAO+CMB data combination), they are derived from our parameter constraints which are quoted with a broadened prior encompassing the full posterior.}
    \label{fig:RE-OCDM}
\end{figure}

As the last of these combinations may at first sight indicate a weak preference for non-zero spatial curvature, we comment on this further in Section \ref{sec:flatness}. We illustrate our results in Figure \ref{fig:RE-OCDM}. 

\subsection{Flat $w$CDM}\label{sec:Cosmo:subsec:wCDM}

\begin{figure}
    \centering
    \includegraphics[width=9cm]{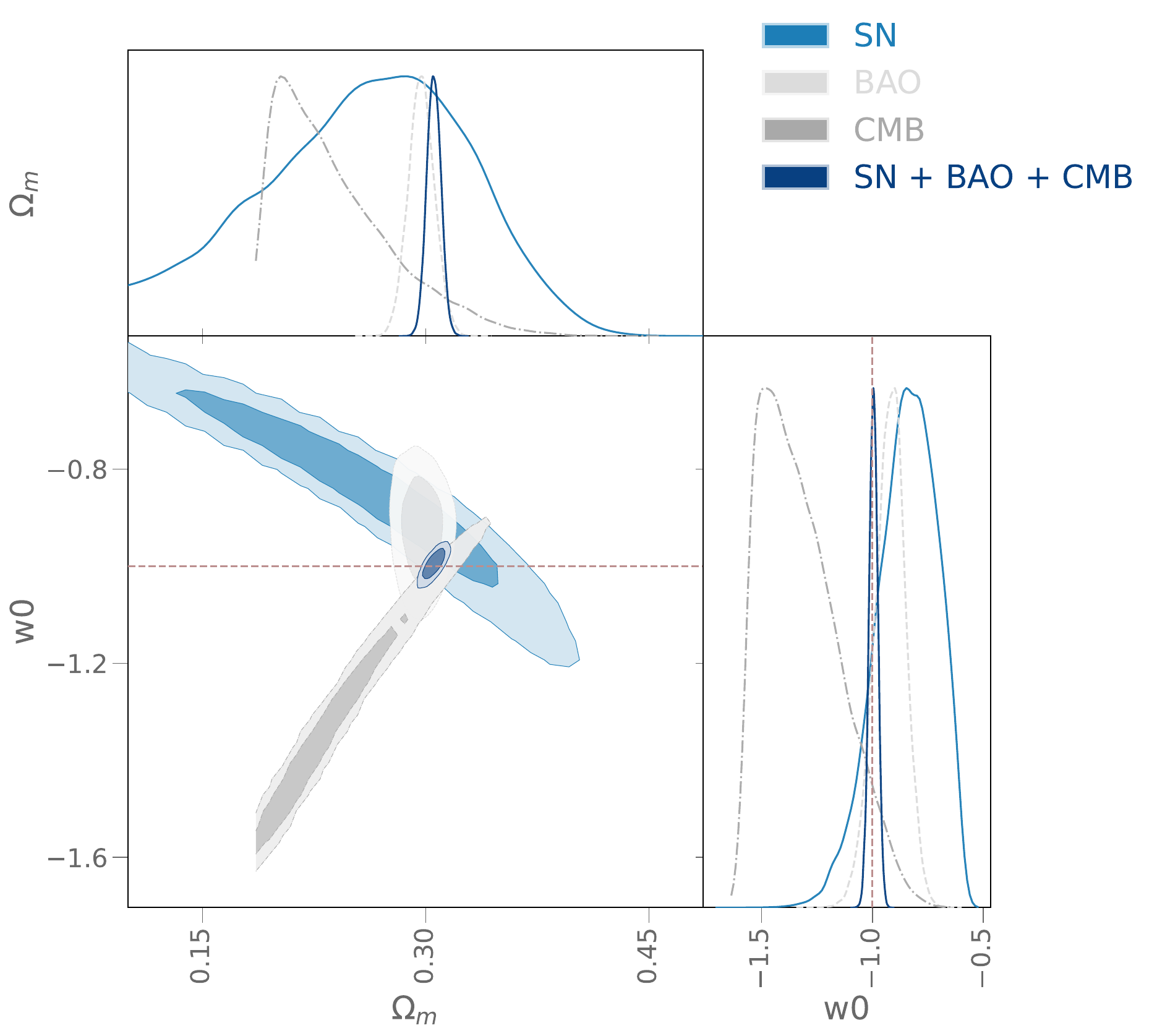}
    \caption{The $\Omega_{\rm m},w$ contours for Flat $w$CDM. In blue, we present SN-only, complemented by the CMB (dark grey) and BAO (light grey) constraints. The full combination of SN+BAO+CMB is in dark blue. We include a maroon dashed line for $w=-1 =$ $\Lambda$CDM.}
    \label{fig:RE-LCDM}
\end{figure}

Fitting DES-Dovekie to Flat-$w$CDM, we find $$\Omega_{\rm m},\; w = 0.263^{+0.064}_{-0.078},\; -0.838^{+0.130}_{-0.142}~(\rm SN~only).$$ DES-Dovekie remains consistent with a cosmological constant at $\sim$1.5$\sigma$, as shown in Figure \ref{fig:RE-LCDM}. As noted in \citet{camilleri2024}, for SN Ia the partial degeneracy between $w$ and $\Omega_{\rm m}$ approximately follows a line of constant deceleration parameter $q_0 = -\ddot{a}a/\dot{a}^2 (z=0)$. For SN, we find $$q_0 = -0.423^{+0.063}_{-0.073}\;.$$

When we combine SN Ia with external probes, we get more precise constraints on our cosmological parameters due to the orthogonal degeneracy directions. Combining with the CMB, we find $\Omega_{\rm m},w = 0.322 \pm 0.007, -0.978 \pm 0.024$ with $\log S = -1.55$ ($1.9\sigma$). When combining with BAO, we find $\Omega_{\rm m},w = 0.297 \pm 0.008, -0.909^{+0.035}_{-0.037}$. 

Finally, with SN+CMB+BAO, we find $$\Omega_{\rm m},\; w = 0.305 \pm 0.005,\; -0.995^{+0.019}_{-0.020}~(\rm SN+BAO+CMB)$$ and $q_0 =  -0.537 \pm 0.026$. While our $w$ constraint might be interpreted that the full data indicates a Universe consistent with a cosmological constant, \citet{Linder2007} has pointed out that values close to $w=-1$ follow from the inclusion of CMB data, almost irrespective of the data at low redshift \citep[see Equations (1)-(3) and following paragraph of text in][]{Linder2007}.{ Furthermore, in this case $\log S = -2.87$ ($2.5\sigma$), indicating the data is not compatible in this model : the tightness of these constraints should be treated with skepticism.}

There is no data combination in which Flat-$w$CDM is preferred to Flat-$\Lambda$CDM in Bayesian evidence.

\subsection{Flat $w_0w_a$CDM}\label{sec:Cosmo:subsec:w0waCDM}

Finally, we present our cosmological fits to Flat $w_0w_a$CDM cosmology. Our SN-only contours are broad with $$w_0, w_a = -0.50^{+0.35}_{-0.27}, -7.5^{+3.6}_{-4.5}~(\rm SN~only),$$ where for this run only, we have enlarged the $w_0w_a$ priors to encompass the full range of significant posterior probability. The wide range of this posterior reflects the degeneracy in the low-redshift Hubble diagram between changes to $\Omega_{\rm m}$ and evolving dark energy. As in $\Lambda$CDM, our constraining power increases as we combine with other probes which help determine $\Omega_{\rm m}$; we find 
$$w_0, w_a = -0.769 \pm 0.100, -0.98^{+0.47}_{-0.49}$$ for SN+CMB, and 
$$w_0, \; w_a =-0.803 \pm 0.054,\; -0.72 \pm 0.21~(\rm SN+CMB+BAO).$$  

{Looking at the pivot redshift, for SN-only we find $z_p = 0.52$ with $w_p = -1.01 \pm 0.13$. It is not entirely expected that we find $w$ close to -1 at the pivot, and we interpret this as re-inforcing the point that DES-Dovekie is entirely compatible with $\Lambda$CDM. For SN+BAO we get $w_p = -0.908 \pm 0.036$, and the suspiciousness statistic $S = -1.86$ ($1.8\sigma$) indicates this combination is reliable. While this hints at a departure from $\Lambda$CDM, the evidence is not very strong. Finally, the combination of SN+BAO+CMB gives $z_p = 0.57$ with $w_p = -0.981 \pm 0.022$ as expected; we emphasize that this is due to compatibility of CMB with $\Lambda$CDM, almost independently of late-time evolution.}

\begin{figure*}
    \centering
    \includegraphics[width=17cm]{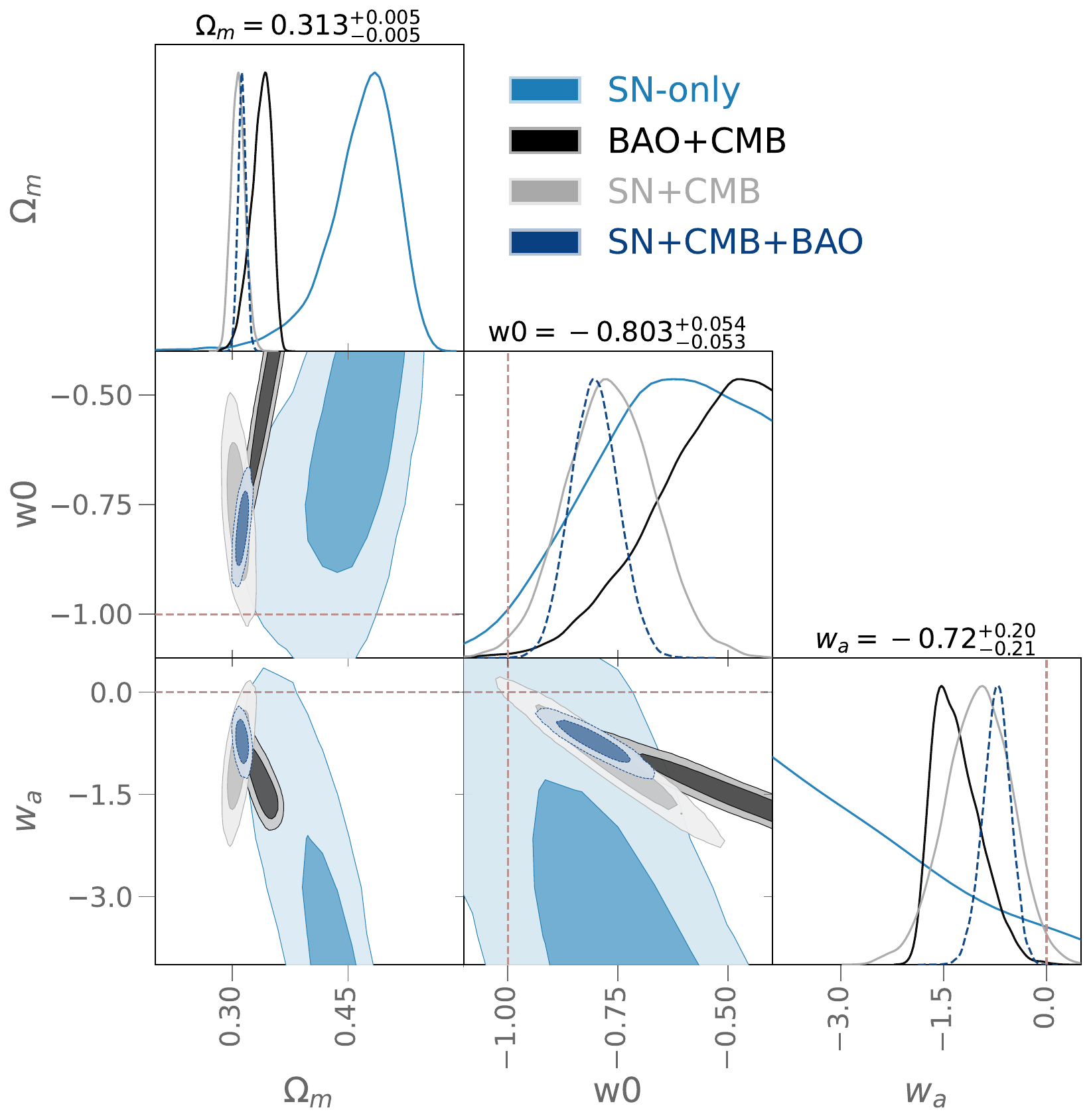}
    \caption{The cosmological contours for $w_0w_a$CDM cosmology. We show SN-only in light blue, SN+CMB in grey, and SN+CMB+BAO in dark blue. Our combined external probes, BAO+CMB, is shown in black. We include light maroon dashed lines for $w=-1 =$ $\Lambda$CDM.} 
    \label{fig:RE-w0wa}
\end{figure*}

We show the contours for SN-only, SN+CMB, and SN+CMB+BAO in Figure \ref{fig:RE-w0wa}. {An interesting feature of broadening the $(w_0,w_a)$ priors for SN-only is that the overlap of the SN and CMB+BAO contours now appears to arise in a low-probability region of the SN contours}\footnote{This probably accounts for why this was not discussed in \citet{DESIDR2}, where the effect must surely have been larger for DES-SN5YR, as only the narrower set of priors were used.}. {Checking the Suspiciousness statistic}, {we find $\log S = -2.02$ ($1.8\sigma$), indicating the data combination \textit{is} reasonable. A further cross-check of the full posterior reveals that large negative values of $w_a$ are associated with $\Omega_{\rm m} > 0.4$ : SN-only data is currently not very good at restricting the extended parameter space of Flat $w_0 w_wa$CDM. Hence, we interpret the visual appearance of the contours as largely a volume effect of the prior. Furthermore, a narrower prior (which would increase compatibility by shifting the posterior up in $w_a$) could be argued for on the basis of constraints on $\Omega_{\rm m}$ arising from data not used in this analysis, for example cosmic shear \citep[e.g.][]{DESY6}.}

Our final cosmological results are summarised in Table \ref{tab:cosmology}, and we discuss the implications for evolving dark energy in Section \ref{sec:evde}. For DES-Dovekie, the number of data points is 1684 which is the effective number of SNe obtained by summing the BEAMS probabilities.

\begin{table}
    \centering
    \begin{tabular}{l|c}
        Parameter & Significance ($\sigma_{\rm syst}$) \\
        \hline
        $w_0$ & $1.1\sigma$ \\
        $w_a$ & $1.2\sigma$
    \end{tabular}
    \caption{Significance of change in cosmological parameter when \textit{only} considering the calibration systematic uncertainty, for SN+CMB+BAO Flat $w_0w_a$CDM cosmology.}
    \label{tab:w0wasyst}
\end{table}

\subsection{Cosmological model preference}

Following our cosmological fits, we assess the preference of the data for models beyond Flat $\Lambda$CDM. As explained in Section \ref{sec:Methodology}, we use two different metrics. One is frequentist \citep[following ][]{DESIDR2}, deriving from the relative probability of the data between two models, and expressed as an equivalent $\sigma$ following Equation \ref{eq:wilk}. The other is the Bayesian evidence \citep[following][]{DES5YRKP}, deriving from the relative probability of two models, given the data. The frequentist metric is positive by construction, whereas for the Bayesian metric a negative value indicates a preference for the extended model compared to Flat $\Lambda$CDM), and a positive value is a preference for Flat $\Lambda$CDM).

For the Bayesian method, we use the interpretative scale given in Table 1 of \citealp{Trotta2008} where $|\Delta \log \mathcal{Z}| < 1$ is inconclusive, $1 < |\Delta \log \mathcal{Z}| < 2.5$ is weak, $2.5 < |\Delta \log \mathcal{Z}| < 5.0$ is moderate and $5.0 < |\Delta \log \mathcal{Z}| < \infty$ is strong evidence for/against the model compared to Flat $\Lambda$CDM. 

For our full data combination, we find Flat $w_0 w_a$CDM is a better fit to the data by $\Delta \chi^2 = -13.5$ versus Flat $\Lambda$CDM, which is a frequentist significance of $3.2\sigma$. However, $\Delta \log \mathcal{Z} = -1.7$, a level which indicates a \textbf{weak} preference for evolving dark energy at odds of $5 : 1$. In this context, there is not a strong evidential basis with the available data to conclude that dark energy evolves.

At first sight, it may appear that the frequentist $3.2\sigma$ and Bayesian $5:1$ model odds are contradictory; we stress that with generally accepted $5\sigma$ threshold for rejection of the null hypothesis (here Flat $\Lambda$CDM), they are consistent in their message. We discuss this further in Section \ref{sec:evde}. 

It is clear from Table \ref{tab:cosmology} that the full combination of data is required to show \textit{any} Bayesian preference for evolving dark energy. We have also tested the BAO+CMB combination, also finding the evidence for Flat $w_0 w_a$CDM is weak. Notably, the SN+CMB combination shows a moderate preference for Flat $\Lambda$CDM compared to Flat $w_0 w_a$CDM. Neither Flat $w$CDM nor $\Lambda$CDM is favoured by any data combination.  

\begin{figure}
    \centering
    \includegraphics[width=9cm]{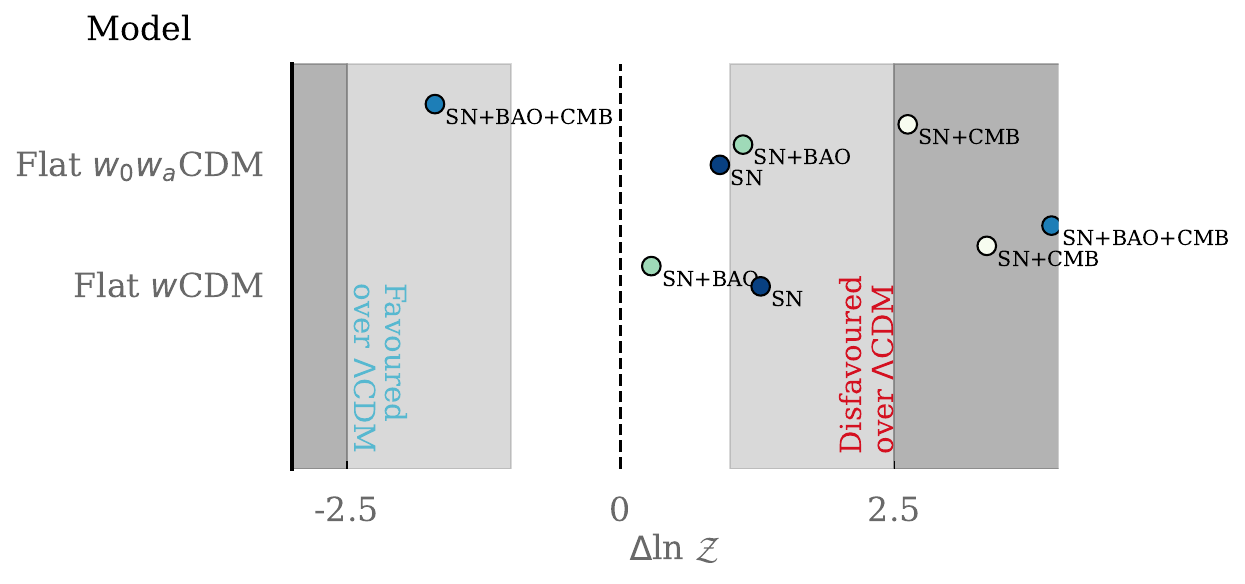}
    \caption{Model preference in terms of the log of the Bayes ratio, $\Delta \log \mathcal{Z}$.}
    \label{fig:BIC}
\end{figure}

Table \ref{tab:tensions} shows a summary of the results, visualised in Figure \ref{fig:BIC}.

\begingroup

\setlength{\tabcolsep}{10pt} 
\renewcommand{\arraystretch}{1.2} 
\begin{table*}
\centering
\caption{Results for our different cosmological models, sorted into sections of different combinations of probes. We present the medians of the marginalised posterior with 68.27\% integrated uncertainties. }
\label{tab:cosmology}
\begin{tabular}{l|cccccrr}
\toprule
\multicolumn{2}{|c|}{$\Omega_{\rm m}$} &$H_0$ & $\Omega_{\rm k}$ & $w_0$ & $w_a$ & $\chi^2$ & $\log \mathcal{Z}$\\ 
\hline
\multicolumn{6}{l|}{\textbf{DES-Dovekie} (SN-only)} \\
\hline
Flat-$\Lambda$CDM & $0.330 \pm 0.015$ & - & - & -  & -  & 1640.3 & -822.5 \\
$\Lambda$CDM      & $0.279 \pm 0.057$& - & $0.14 \pm 0.15$ & -  & -  & 1639.5 & -822.5 \\
Flat-$w$CDM       & $0.263^{+0.064}_{-0.078}$ & - & - & $-0.838^{+0.130}_{-0.142}$ & -  & 1639.0 & -823.8 \\
Flat-$w_0w_a$CDM  & $0.473^{+0.035}_{-0.050}$ & - & - & $-0.497^{+0.348}_{-0.267}$ & $-7.46^{+3.60}_{-4.48}$ & 1634.2 & -823.4 \\
\midrule
\multicolumn{6}{l|}{\textbf{DES-Dovekie} + CMB} \\
\hline
Flat-$\Lambda$CDM & $0.317 \pm 0.005$ & $67.29 \pm 0.34$ & - & -  & -  & 2224.1 & -1144.4 \\
$\Lambda$CDM      & $0.335 \pm 0.012$ & $65.04 \pm 1.24$ & $-0.0063 \pm +0.0037$  & - & - & 2219.9 & -1145.4 \\
Flat-$w$CDM       & $0.322 \pm 0.008$ & $66.70 \pm 0.71$ & - & $-0.978 \pm 0.024$ & -  & 2223.1 & -1147.7\\
Flat-$w_0w_a$CDM  & $0.308 \pm 0.009$ & $68.11 \pm 0.89$ & - & $-0.769 \pm 0.100$ & $-0.98 \pm 0.48$ & 2219.5 & -1147.0 \\
\midrule
\multicolumn{6}{l|}{\textbf{DES-Dovekie} + BAO} \\
\hline
Flat-$\Lambda$CDM & $0.306 \pm 0.008$  & - & - & -  & -  & 1654.5 & -833.2 \\
$\Lambda$CDM      & $0.293 \pm 0.011$ & - & $0.054^{+0.033}_{-0.036}$  & -  & - & 1652.4 & -833.3 \\
Flat-$w$CDM       & $0.297 \pm 0.008$ & - & - & $-0.909^{+0.035}_{-0.037}$ & -  & 1648.6 & -833.5\\
Flat-$w_0w_a$CDM  & $0.313^{+0.013}_{-0.016}$ & - & - & $-0.843^{+0.071}_{-0.065}$ & $-0.53 \pm 0.44$ & 1647.2 & -834.3\\
\midrule
\multicolumn{6}{l|}{\textbf{DES-Dovekie} + CMB + BAO} \\
\hline
Flat-$\Lambda$CDM & $0.304 \pm 0.003$ & $68.14 \pm 0.23$ & - & -  & -  & 2244.0 & -1155.6\\
$\Lambda$CDM      & $0.305 \pm 0.003$ & $68.57 \pm 0.30$ & $0.0026 \pm 0.0011$ & -  & -  & 2238.4 & -1156.7 \\
Flat-$w$CDM       & $0.305 \pm 0.005$ & $68.02 \pm 0.53$ & - & $-0.995 \pm 0.019$ & -  & 2244.5 & -1159.6\\
Flat-$w_0w_a$CDM  & $0.313 \pm 0.005$ & $67.47 \pm 0.55$ & - & $-0.803 \pm 0.054$ & $-0.72 \pm 0.21$ & 2230.5 & -1153.9 \\
\midrule

\bottomrule
\end{tabular}

\end{table*}

\endgroup

\begin{table*}
    \centering
    \begin{tabular}{l|cc|cc}
        \textbf{Model}  & $\Delta\chi^2_{\rm ML}$ & $\sigma$ &  $\Delta\log \mathcal{Z}$ & Bayesian model preference\\
        \hline
        Open $\Lambda$CDM & -5.6 & 2.3$\sigma$ & +1.1 & Weakly not preferred \\
        Flat $w$CDM & 0.0 & 0.0$\sigma$ & +4.0 & Moderately not preferred\\ 
        Flat $w_0w_a$CDM & -13.5 & \textbf{3.2$\sigma$}& -1.7 & \textbf{Weakly preferred}
    \end{tabular}
    \caption{The model preference metrics for our most constraining data combination, DES-Dovekie + CMB + BAO compared to Flat $\Lambda$CDM. We provide the conversion of the $\Delta \chi^2_{\rm ML}$ into equivalent $\sigma$ using Wilk's theorem and Eqn. \ref{eq:wilk}. The logarithm of the Bayes ratio is interpreted using an adaptation of the Jeffreys scale, given in Table 1 of \citet{Trotta2008}, where $|\Delta \log \mathcal{Z}| \in (0,1), (1,2.5), (2.5, 5.0), (5.0, \infty)$ is inconclusive, weak, moderate and strong evidence respectively. A positive sign indicates the model is not preferred compared to Flat $\Lambda$CDM, and a negative sign means it is preferred.}
    \label{tab:tensions}
\end{table*}

\section{Discussion and Conclusions}\label{sec:Conclusions}

\subsection{Comparison to DESI DR2}

We benchmark our analysis choices by combining the original DES-SN5YR dataset \citep[as used in][]{DESIDR2} with BAO, but with our updated CMB data now including ACT and SPT (which were not available to DESI at their time of writing, but see \citet{GarciaQuintero2025} for a discussion of the addition of ACT). We find a frequentist $4.0 \sigma$ ($\Delta \chi^2 = -19.2$) compared to $4.2 \sigma$ ($\Delta \chi^2 = -21.0$) in DESI DR2 (\citet{DESIDR2} does not quote Bayesian evidences). The alignment is in part coincidental as the CMB likelihoods differ, and this is discussed in Section \ref{sec:evde}. We have also made use of the {Nautilus} sampler in {Cosmosis} rather than {Cobaya} MCMC sampler, but since our chains are well-converged we expect this to be a minor effect. Therefore, our results for DES-SN5YR are consistent with \citet{DESIDR2} within the bounds of our differing choices. 

\subsection{The evidence for spatial flatness}
\label{sec:flatness}
It is well-known that Planck PR3 data prefers a non-spatially flat Universe at over $2\sigma$ preference, however once low-redshift lensing, Pantheon+ supernovae or BAO data from SDSS is included the preference disappears (see Section 7.3 in \citet{Planck20}, and \citet{Efstathiou2020}). Interestingly, the over $2\sigma$ preference is restored when the CMB is combined with DESI DR2 BAO data (Tables V and VI of \citealp{DESIDR2}). As noted by \citet{Chen2025}, a non-spatially flat model provides a viable alternative to evolving dark energy when considering solely this data combination. For our data combination of SN+BAO+CMB, we find $\Omega_{\rm k} = 0.0026 \pm 0.0011$, apparently also indicating a small non-zero spatial curvature. 

However, we note the data combination of SN+CMB has $ \Omega_{\rm k} = -0.0063 \pm 0.0037$ (recall our CMB combination includes lensing data from Planck, ACT and SPT) whereas for BAO+CMB it is $ \Omega_{\rm k} = 0.0025 \pm 0.0011$. The positive non-zero $\Omega_{\rm k}$ for BAO+CMB appears to be driven by DESI DR2 favouring a lower $\Omega_{\rm m}$ than the CMB, which drives a slight internal tension with SN Ia in $\Lambda$CDM. This is borne out in the SN+BAO+CMB combination, as the $\Lambda$CDM model is weakly disfavoured relative to Flat-$\Lambda$CDM and moderately disfavoured relative to Flat-$w_0 w_a$CDM. 
 
Therefore, we conclude there is no evidence for non-zero spatial curvature arising from DES-Dovekie, DESI DR2 and CMB data. 

\subsection{The Hubble constant}
\label{sec:hubble}
The tension between the Hubble constant as measured from the local distance ladder, and as fit by a full redshift range of cosmological data remains an enduring mystery. For the CMB, \citet{Planck20} reported $H_0 = 67.36 \pm 0.54$ km sec$^{-1}$ Mpc$^{-1}$, ACT reports $66.11 \pm 0.79$ km sec$^{-1}$ Mpc$^{-1}$ \citet{ActDr6like}, and SPT-3G gave $H_0 = 66.66 \pm  0.60$ km sec$^{-1}$ Mpc$^{-1}$ \citet{Camphuis2025}. For the local distance ladder, recent results include $H_0 = 73.29 \pm 0.90$ km sec$^{-1}$ Mpc$^{-1}$ \citep{Murakami2023} and $H_0 = 70.39 \pm 1.94$ km sec$^{-1}$ Mpc$^{-1}$ \citep{Freedman2025}, with large differences in the estimation of systematics. Improved photometry from the James Webb Space telescope has not revealed any errors with earlier estimates due to lower resolution Hubble Space Telescope photometry (see e.g. \citealp{Riess2025}), and differences in local distance ladder may be traced to sample differences \citep{Riess2024}. Alternative probes have not matured as quickly as previously anticipated to provide an arbiter between these two clusters of results that is convincingly free of systematics \citep[for a review, see e.g.][]{Shah2021}.

Of course, BAO and SN Ia by themselves do not constrain $H_0$ as only relative observations are taken. The CMB constrains $H_0$ through the detailed shape of its power spectrum, which constrains the matter, baryon and photon densities prior to recombination, and hence the absolute scale of the temperature fluctuations (at least, in models without exotic additional physics pre-recombination). Combined with the angular size, the distance to the surface of last scattering is well-constrained. Given a model for the subsequent evolution (by using the matter density and dark energy evolution as constrained by the CMB and low redshift data, and again assuming no further exotic physics), $H_0$ is constrained \citep[for a review, see e.g.][]{Lemos2024}.

In Flat-$\Lambda$CDM, our constraint of $H_0 = 68.14 \pm 0.25$ km sec$^{-1}$ Mpc$^{-1}$ is discrepant from \citet{Murakami2023} at the level of $5.5\sigma$. However, the errors are primarily driven by the CMB and BAO data rather than SN Ia for this model. It is modestly tighter than $H_0 = 67.24 \pm 0.35$ km sec$^{-1}$ Mpc$^{-1}$ quoted in Table I of \citet{Camphuis2025} for their CMB-SPA combination, due to the inclusion of BAO data. Similarly, is also modestly tighter than $68.17 \pm 0.28$ km sec$^{-1}$ Mpc$^{-1}$ quoted in Table V of \citet{DESIDR2} for their DESI+CMB combination, due to the inclusion of more CMB data. 

Additional variation of $H_0$ is allowed in extended models of dark energy. The general trend when adding DESI DR2 data is for lower $H_0$ \citep[as pointed out by][]{Lewis2025} which works against the larger error bars in converting to a tension. For the extended model, it is notable that SN Ia data \textit{is} helpful in constraining $H_0$. The CMB+BAO combination results in $64.88^{+1.52}_{-1.06}$km sec$^{-1}$ Mpc$^{-1}$ whereas CMB+SN Ia has $68.11 \pm 0.89$km sec$^{-1}$ Mpc$^{-1}$, a discrepancy of $1.8\sigma$. Interestingly, our combined CMB+BAO+SN Ia constraint for Flat-$w_0 w_a$CDM is $H_0 = 67.47 \pm 0.55$ km sec$^{-1}$ Mpc$^{-1}$ is within $0.2\sigma$ of the Flat $\Lambda$CDM value for CMB data only. 

Our combination of CMB+BAO+SN Ia represents an inverse distance ladder. In effect, the matter density of the Universe throughout cosmic time and the absolute size of the BAO are determined by the CMB (and the pre-CMB evolution is assumed to be free of exotic physics such as Early Dark Energy). The residual departure of SNe and BAO distances in the late Universe are small, and able to be captured within the framework of Flat $w_0 w_a$CDM \citep[and more general evolution, as emphasised in][]{Lodha2025}. The question of whether dark energy evolves or not in the late Universe is unconnected with any potential resolution of the Hubble constant tension. 

\subsection{The evidence for evolving dark energy}
\label{sec:evde}

In this paper, we have presented a re-analysis of the DES-SN5YR data with an updated filter recalibration, incorporating extra degrees of model freedom and more calibration data. The result of our re-analysis is that the evidence for Flat $w_0 w_a$CDM over Flat $\Lambda$CDM is reduced by $\Delta \log \mathcal{Z}$ $3.5$ compared to DES-SN5YR, which is enough to downgrade the preference for evolving dark energy from strong to weak.

As discussed previously, model preference can be assessed both in a frequentist or Bayesian framework. While a detailed review is beyond the scope of this paper, we should expect that differing choices be consistent in their message. We consider a frequentist $3.2\sigma$ to be consistent with our classification of the log Bayes ratio of $-1.7$ (model odds of 5:1) as weak evidence in favour of evolving dark energy, as it is generally accepted that $5\sigma$ is an appropriate threshold for a strong rejection of the null hypothesis that the Universe is Flat $\Lambda$CDM. In a hypothetical scenario where SN+CMB+BAO data indeed indicated a frequentist preference of $5\sigma$, we estimate the log Bayes ratio would be $\sim -8$, a level which would be indeed be classified as strong on our interpretative scale. While Bayesian evidence does depend on the choice of prior, within reasonable ranges of choices (that is, broadly consistent with other astrophysical and physical constraints, such as structure formation) this equivalence would not be materially affected.

In terms of the Hubble diagram, our change in preference for evolving dark energy can be attributed to the reduction in the relative dimness of low-redshift SNe, relative to mid-redshift SNe, as is apparent in Figure \ref{fig:HUBBLE}. This in turn is driven by a change in the SNe light curve fitting model SALT (which is trained on data from multiple surveys and hence multiple filters) and a change in the effective definition of the DES$-g$ band, from \citetalias{Dovekie} restoring the original PS1$-g$ reference filter, as shown in Figure 14 in \citetalias{Dovekie}. We emphasize that this change is driven solely by improvement to the data (in particular, the addition of DA white dwarfs as calibrator stars) and methodology used for filter recalibration, and corrections to the shape of the colour law. It is also consistent with previously reported systematics. Table \ref{tab:w0wasyst} shows that our $\sigma(\rm phot)_{\rm syst}$ uncertainties are well-calibrated.

The preference for $w_0w_a$CDM cosmology is not driven by changes to the SNe alone. The influence of the choice of low-$\ell$ CMB likelihood on the preference for evolving dark energy has been noted in \citet{Sailer2025, GarciaQuintero2025}. In general terms, $\Omega_{\rm m}$ determines the peak heights of the CMB power spectrum, but also so does the Thomson scattering of CMB photons by the charged inter-galactic medium in the epoch post-reionisation. This leads to a degeneracy between $\Omega_{\rm m}$ and the optical depth parameter $\tau$, which is determined by the CMB polarisation. Higher $\tau$ leads to a lower preference for evolving dark energy. We have tested the impact of different choices of CMB likelihood by successively removing SPT and ACT data, and replacing the low-$\ell$ \texttt{simall} likelihood with a $\tau$-prior from the \texttt{Sroll2} likelihood \citet{Delouis2019} (as is done in \citealp{ActDr6like}). The choices all \textit{reduce} our preference for evolving dark energy by between 0.2 to 0.7$\sigma$, or in Bayesian terms by up to $\delta \log \mathcal{Z} \sim 3.0$. Although we do not test specifically the combination used in DESI DR2, it is likely given the low optical depth noted for that choice in \citet{rosenberg2022} that the preference for evolving dark energy from that choice is at the upper end of the combinations we tested.

All told, our results place DES-Dovekie as intermediate in preference for evolving dark energy, in between Pantheon+ and Union3 (which \citet{DESIDR2} quotes as $\Delta \chi^2 = -10.7$ and $-17.4$ respectively), and consistent with both. Moreover, it should be emphasised that the three supernova data sets in common use : Pantheon+, Union3, and DES-Dovekie are all consistent with each other, with differences between them at the $\lesssim 1\sigma$ level in terms of cosmological parameters. Approximately adjusting these numbers for our analysis choices, we estimate that a Bayes ratio for the Pantheon+ dataset would also show no evidence in favour of $w_0 w_a$CDM, and Union3 would moderately prefer it, indicating that our conclusion that there is no strong preference for evolving dark energy is robust to other choices of SN Ia data.

\subsection{Improvements over DES-SN5YR}

This work is best understood as part of an ongoing effort within the SN\,Ia community to improve modelling and facilitate reanalysis of photometric data. This continues the trend of greater data volumes informing the modelling techniques, in turn leading to better understanding of systematics, that has been in progress over the last 25 years. 

It has been argued that the evidence for evolving dark energy reported in \citet{DESIDR2} is a consequence of SNe systematics \citep{Efstathiou2024} or inconsistencies between cosmological parameters across probes \citep{Tang25}. While the apparent inconsistency of SN Ia distances in common to different datasets was largely explained as the consequence of differing selection functions, scatter models, and mass calibration in \citet{Vincenzi2025}, the question remains : how safe are SNe data?

Broadly, most supernovae systematics fall into the three categories of photometric calibration, foregrounds or astrophysical processes. Our paper addresses the first of these, and may be regarded as a continuation of the investigation of \citet{Vincenzi2025}. 

During our re-creation of the DES-SN5YR dataset, we discovered an outdated approximation used for the \citet{Fitzpatrick99} colour law. We also found that the systematic uncertainty for calibration had been under-weighted (Section \ref{sec:Differences}), and this was corrected here as well. Considering sequential changes from DES-SN5YR, the colour law corrected DES-SN5YR increases $\Omega_{\rm m}$ by $\sim 1 \sigma$, and then DES-Dovekie photometric recalibration decreases $\Omega_{\rm m}$ by $\sim 2 \sigma$. Although these changes are notable when considered in isolation, this neglects the correlation with the rest of the pipeline, and the change is less significant in extended models (Table \ref{tab:syst_size}). 

The new calibration and updated SALT model do not cause any obvious changes to our observed distribution of SN Ia properties, save that of the DES $c$ distribution, which is different at $2.8\sigma$ from DES-SN5YR. Investigating further, we find that this change in the DES $c$ distribution is predicted when simulating with the \citetalias{DES5YR} simulation parameters, resulting in good agreement between our simulated and observed $c$ distributions (Figure \ref{fig:DATASIM}). Therefore, the change in our observed colour is entirely explained by the change of SALT model.

It is a familiar refrain that the $\sim200$ SNe low-redshift sample in DES-SN5YR (and earlier datasets) requires modernising. Unfortunately, some of these older low-redshift samples, such as CfA3, contain spectral sequences that are believed essential for SALT training. Incorporating these spectral sequences involves cross-calibrating these older samples, thereby increasing the calibration systematic even in the case of single-telescope cosmology analyses (e.g. a DECam low-$z$ sample+DES, or PS1+Foundation). Fully replacing these older samples will require either more complete knowledge of the impact of dropping these spectral sequences, or a sample that includes these sequences with a better-understood calibration path. 

Given the change in the SALT model, calibration, \textit{and} the F99 colour law, a natural question that arises is the necessity of recalibrating the scatter model by updating the Dust2Dust fitting code \citep{Popovic22} to use the exact F99 colour law. We find that the $\chi^2/{\nu}$ changes by $<1$ between the approximate and exact F99 colour laws, which we did not find sufficient to require a re-determination of the intrinsic SN Ia and dust parameters. When comparing the Dust2Dust metrics from the original SALT3.DES5YR ($c, \mu_{\rm res}, \sigma\mu_{\rm res}$), we find that the new data/simulation agreement is slightly worse than the original \citetalias{DES5YR}. This behaviour is not driven by any individual metric, but rather a general trend.  Given recent improvements in our understanding of host galaxy environs \citep{Gonzalez-Gaitan21, Kelsey23, Grayling24, Ginolin24b}, there is ample opportunity for future work to improve upon dust modelling. Upcoming low-redshift samples such as ATLAS \citep{ATLASSURVEY} or DEBASS \citep{Acevedo25, Sherman25} in the near future, or ZTF \citep{Rigault25} and LS4 \citep{LS4} present exciting opportunities for combination with DES-Dovekie. 

{The LSST-TiDES \citep{Frohmaier2025} survey is forecast to assemble a Hubble diagram of $\sim 140,000$ SN Ia with spectroscopic redshifts (either directly from the SN Ia spectral sequence, or the host galaxy). On current systematic errors therefore, this data would be systematic-limited. Fortunately there are grounds for optimism that systematics will keep pace, with the above mentioned improvements in low-z data, plus LSST-TiDES data itself. Low-z SN Ia are useful not only to control systematics : the Universe starts to accelerate at $z \sim 0.7$, and becomes dark energy dominated at $z \sim 0.3$. While high-z SN Ia are important to establish consistency of the matter density between the CMB, BAO and SN Ia, it is in the redshift range $z<0.5$ that we obtain most information about potential dark energy evolution. }

In summary, DES supernovae remain the best-characterised high-redshift sample before LSST. Further improvements to the understanding of environmental factors and their impact on cosmology will doubtless be achieved; but DES-Dovekie demonstrates that the source of the current $\Lambda$CDM tension is very unlikely to arise solely from photometric cross-calibration. As such, we recommend the distances in this paper supersede the original DES-SN5YR distances.

\section*{Acknowledgements}

This work was completed in part with resources provided by the University of Chicago's Research Computing Center. This project has received funding from the European Union’s Horizon Europe research and innovation programme under the Marie Skłodowska-Curie grant agreement No 101205780. This work has been supported by the research project grant “Understanding the Dynamic Universe” funded by the Knut and Alice Wallenberg Foundation under Dnr KAW 2018.0067 and the {\em Vetenskapsr\aa det}, the Swedish Research Council, project 2020-03444.

B.P. acknowledges you, gentle reader. P.W. acknowledges support from the Science and Technology Facilities Council (STFC) grant ST/Z510269/1. T.D. acknowledges support from the Australian Research Council through the ARC Centre of Excellence for Gravitational Wave Discovery (OzGrav), CE230100016. LK acknowledges support for an Early Career Fellowship from the Leverhulme Trust through grant ECF-2024-054 and the Isaac Newton Trust through grant 24.08(w). A.M. is supported by the Australian Research Council DE230100055. P.A. acknowledges that parts of this research was carried out on the traditional lands of the Ngunnawal people. We pay our respects to their elders past, present, and emerging. L.G. acknowledges financial support from AGAUR, CSIC, MCIN and AEI 10.13039/501100011033 under projects PID2023-151307NB-I00, PIE 20215AT016, CEX2020-001058-M, ILINK23001, COOPB2304, and 2021-SGR-01270.


\section{Author Contributions}

B.P. performed the analysis, drafted the manuscript, and determined the visual language of the work. P.S. contributed to the analysis -- cosmology fits and statistical tests for cosmological preferences in particular -- and  manuscript preparation. W.D.K. aided in the analysis, unblinding, and writing of the paper. R.K. performed continuation and upgrades of {SNANA} and general writing/issue-solving aid. T.D., provided detailed feedback on the analysis, and along with D.S., served as internal reviewer. A.G. provided writing and analysis guidance. M.V. provided discussion on cosmological results and pipeline implementation. P.W., R.C., E.C. provided help to generate simulations. J.F., L.G., J.L., A.M., and B.S. provided comments on the manuscript.
The remaining authors have made contributions to this paper that include, but are not limited to, the construction of DECam and other aspects of collecting the data; data processing and calibration; developing broadly used methods, codes, and simulations; running the pipelines and validation tests; and promoting the science analysis. 

\section{Data Availability}

Data used in this article is publicly available with the DES-SN5YR data release from \cite{Sanchez24}, hosted at https://github.com/des-science/DES-SN5YR.

The analysis files used for this work are included with the public SNANA install on Zenodo, and the data products of this work are available publicly at https://github.com/des-science/DES-SN5YR

\appendix
\section{Fitzpatrick 99 Changes}\label{sec:F99}

\begin{figure}
    \centering
    \includegraphics[width=8cm]{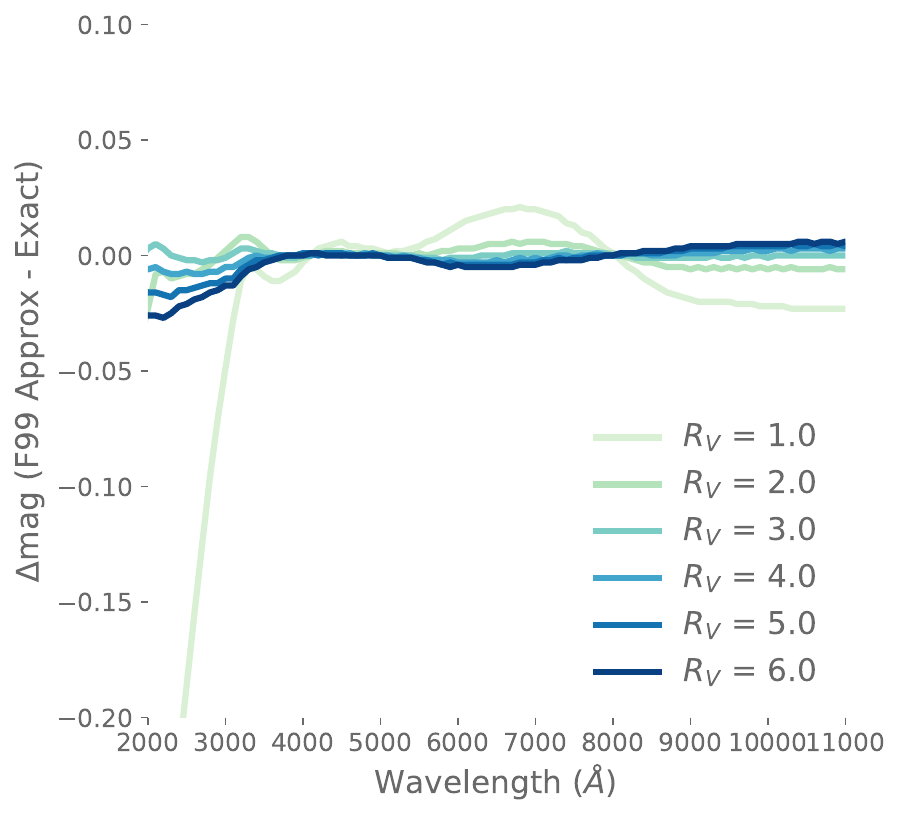}
    \caption{The magnitude difference between the former F99 approximation and the exact F99 colour law, as a function of wavelength. We show this for several selected $R_V$ values, from $R_V = 1$ to $R_V=6$. }
    \label{fig:F99plot}
\end{figure}

Since approximately 2013, {SNANA} had used a polynomial expansion of the \cite{Fitzpatrick99} colour law, centred at $R_v = 3.1$. In the process of integrating BayeSN \citep{Grayling24, Thorp24, Mandel22} into SNANA, this approximation was discovered by comparing the original BayeSN code with the {SNANA} development version. The {SNANA} polynomial was replaced with the full F99 colour law by Dr. Stephen Thorp in early 2025. In Figure \ref{fig:F99plot}, we show the $\Delta$mag as a function of wavelength between the previous approximation and the exact colour law, for various $R_V$ values. 

For the $R_V \geq 2$ regime, these changes are close to negligible within the SALT model range (2000-11000 \r{A}). However, \citetalias{DES5YR}, in addition to Amalgame \citep{Popovic24a} and Pantheon+, use simulated supernovae drawn from a Gaussian distribution that spans approximately $1.4 < R_V \leq 4$, necessitating a regeneration of the bias correction simulations. The impact of this change is shown in Appendix \ref{sec:F99Cosmo}.

\section{Cosmology with F99 Fix}\label{sec:F99Cosmo}

\begin{figure}
    \centering
    \includegraphics[width=8cm]{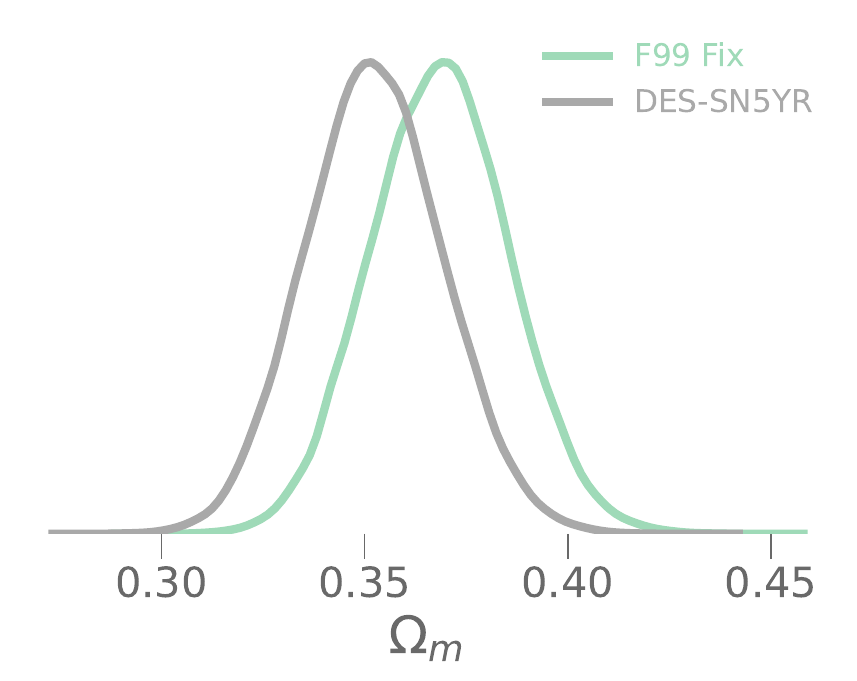}
    \caption{$\Omega_{\rm m}$ posteriors in Flat-$\Lambda$CDM for the original DES-SN5YR (grey) and the F99 colour law fix (light green). The F99 Fix moves the matter density \textit{before re-calibration} to be relatively higher compared to the CMB by $\sim 1 \sigma$.}
    \label{fig:F99lcdm}
\end{figure}

\begin{figure*}
    \centering
    \includegraphics[width=17cm]{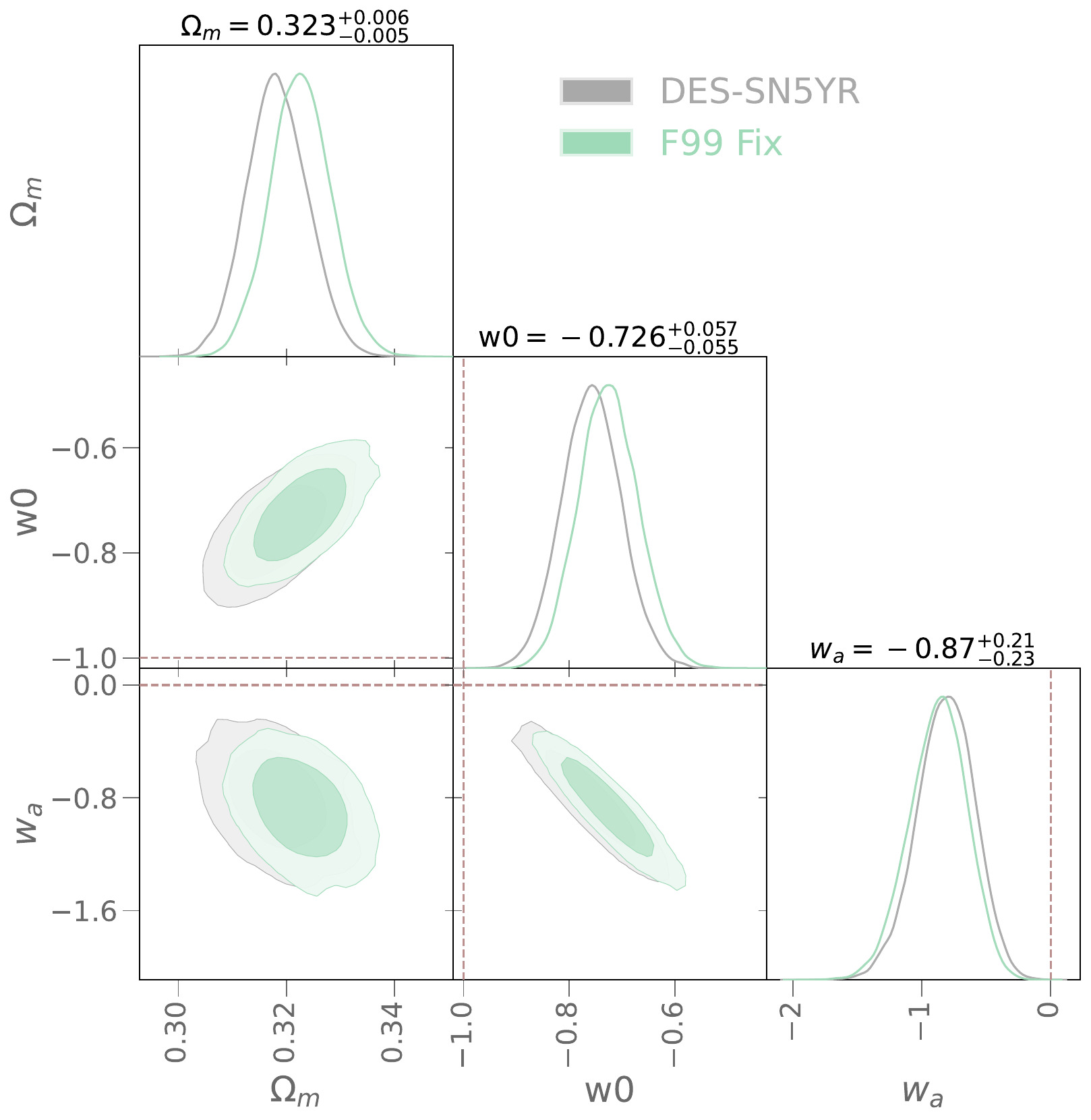}
    \caption{$\Omega_{\rm m}$, $w_0$, and $w_a$ posteriors, from the combined SN+BAO+CMB sample for DES-SN5YR (grey) and the F99 colour law fix (light green). The F99 fix increases the discrepancy with $\Lambda$CDM (note this is \textit{before re-calibration}).}
    \label{fig:F99w0wa}
\end{figure*}

Figures \ref{fig:F99lcdm} and \ref{fig:F99w0wa} shows our posteriors for Flat $\Lambda$CDM and $w_0w_a$CDM cosmologies, after updating the \citet{DES5YR} pipeline to include the fix to the F99 colour law, \textit{but before the Dovekie recalibration is done}. In $\Lambda$CDM we find $\Omega_{\rm m} = 0.369 \pm 0.017$, a level which is $2.8\sigma$ discrepant with the Planck CMB. In $w_0 w_a$CDM we find $w_0, w_a = -0.74 \pm 0.06, -0.80^{+0.23}_{-0.24}$, increasing the inconsistency of the data with $\Lambda$CDM by $0.3\sigma$.

\section{Comparison to DES-SN5YR}\label{sec:AdditionalCosmo}

\begin{table}
    \centering
    \begin{tabular}{l|ccc}
        Model & $\Delta \Omega_{\rm m}$ & $\Delta w$ & $\Delta w_a$ \\
        \hline
        SN-Only Flat $\Lambda$CDM & -0.022 & -- & -- \\
        SN-Only $\Lambda$CDM & -0.020 & -- & -- \\
        SN-Only Flat $w$CDM & -0.007  & -0.031 & -- \\  
        SN-Only Flat $w_0w_a$CDM & -0.023 & -0.138  & 1.47 \\
         & & & \\
        SN+CMB+BAO Flat $\Lambda$CDM & -0.001 & -- & -- \\
        SN+CMB+BAO $\Lambda$CDM & -0.001  & -- & -- \\
        SN+CMB+BAO Flat $w$CDM & -0.004 & -0.015 & -- \\
        SN+CMB+BAO Flat $w_0w_a$CDM & -0.006 & -0.052 & 0.130 \\
    \end{tabular}
    \caption{A selection of the DES-Dovekie - DES-SN5YR changes in cosmology.}
    \label{tab:cosmodiff}
\end{table}

\begin{figure}
    \centering
    \includegraphics[width=9cm]{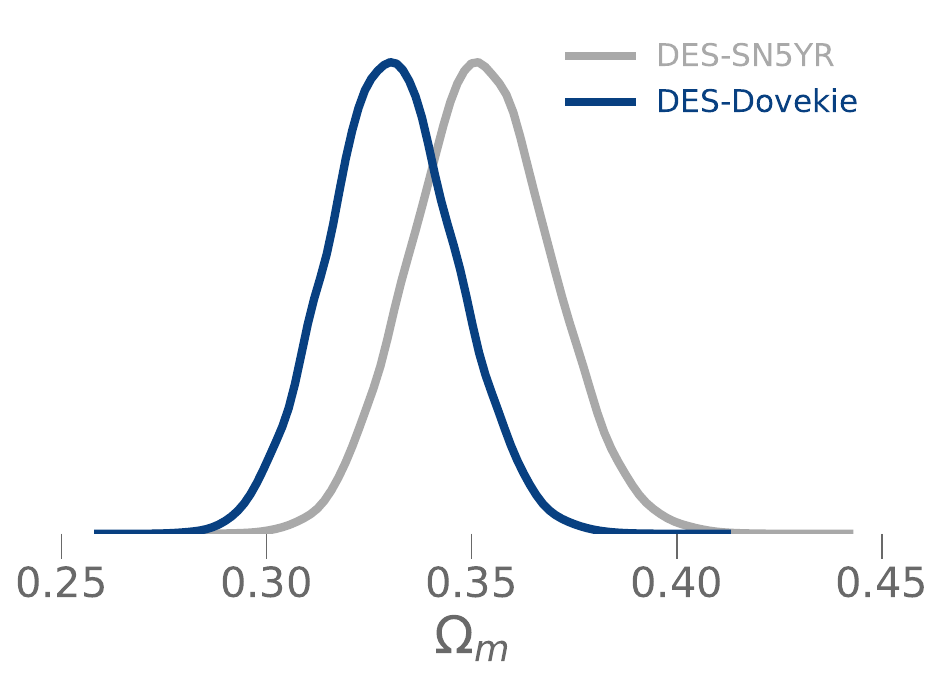}
    \caption{$\Omega_{\rm m}$ posteriors for Flat $\Lambda$CDM for the original DES-SN5YR (grey) and DES-Dovekie (dark blue).}
    \label{fig:flcdmDIFF}
\end{figure}

\begin{figure}
    \centering
    \includegraphics[width=9cm]{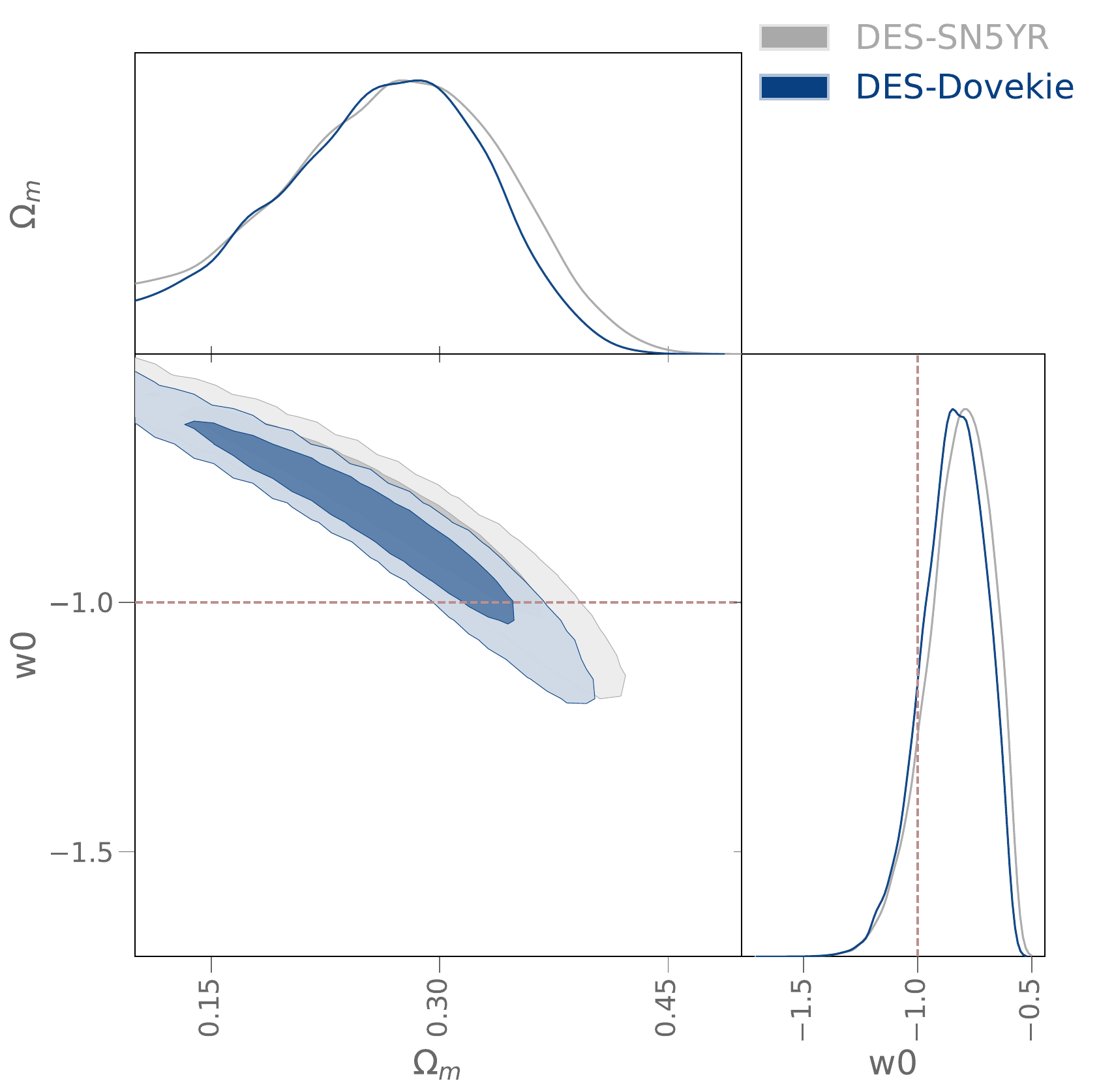}
    \caption{Comparison between DES-SN5YR and DES-Dovekie contours for the Flat-wCDM model. In grey, we present the DES-SN5YR contours, and the DES-Dovekie contours in dark blue. We provide a light maroon dashed line at $w=-1$, the cosmological constant. }
    \label{fig:wCDM-compare}
\end{figure}

In this section, we highlight the changes in cosmology results between DES-SN5YR and DES-Dovekie. Table \ref{tab:cosmodiff} shows the difference in cosmological results for SN-only and SN+CMB+BAO, the most instructive and most interesting combination of probes respectively. Figure \ref{fig:flcdmDIFF} shows the SN-only Flat $\Lambda$CDM posterior for DES-Dovekie and DES-SN5YR, and Figure \ref{fig:wCDM-compare} shows the same but for $w$CDM. 

\begin{figure*}
    \centering
    \includegraphics[width=17cm]{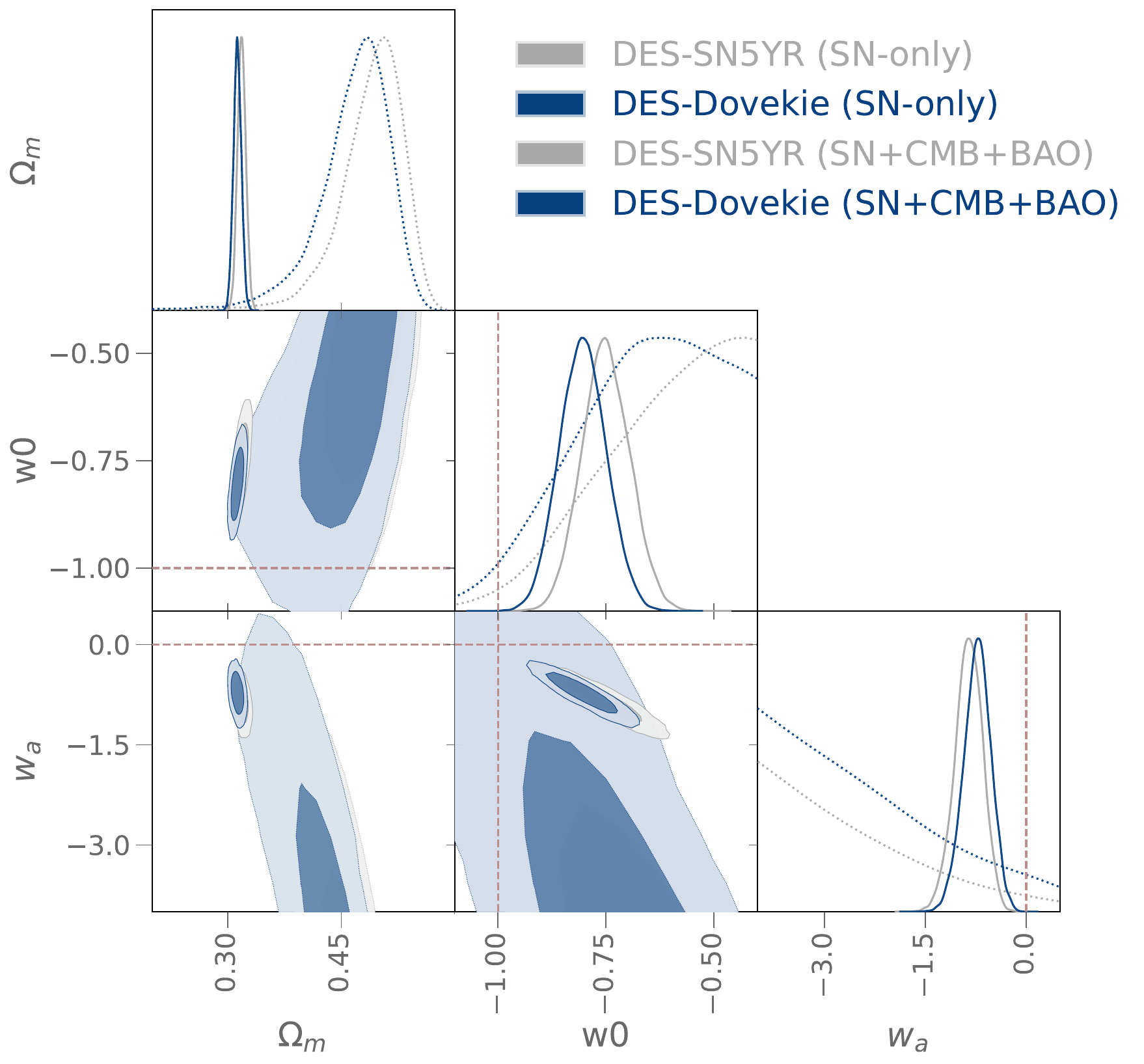}
    \caption{$\Omega_{\rm m}$, $w_0$, and $w_a$ posteriors for SN+CMB+BAO with SN Ia as DES-SN5YR (grey) and DES-Dovekie (dark blue). We present the SN-only posteriors for each in high-transparency.  }
    \label{fig:w0waDIFF}
\end{figure*}

In all cases, DES-Dovekie finds a lower $\Omega_{\rm m}$, resulting in greater consistency between SN, CMB and BAO. For $w_0$ and $w_a$, DES-Dovekie prefers $w_0$ closer to $-1$ and $w_a$ slightly closer to zero. The shifts in $w$CDM are within our systematic uncertainty for calibration as in Table \ref{tab:systematic_uncertainties}, a qualitative $0.5\sigma$. Figure \ref{fig:w0waDIFF} compares the SN-only constraints for DES-SN5YR and DES-Dovekie. 

DES-Dovekie parameter uncertainties are lower compared to DES-SN5YR, as a result of the greater internal consistency of the data. $\sigma(\Omega_{\rm m})$ is 10\% lower in Flat $\Lambda$CDM, Flat $w$CDM and Flat $w_0 w_a$CDM, as is $\sigma(w_0)$ and $\sigma(w_a)$ in extended dark energy models. 

Finally, we note that the d.o.f. of DES-Dovekie are 1684, slightly lower than DES5YR. This value is obtained by summing the BEAMS probability from BBC.

\section{Affiliations}

$^1${Universite Claude Bernard Lyon 1, CNRS, IP2I Lyon / IN2P3, IMR 5822, F-69622 Villeurbanne, France}

$^2${School of Physics and Astronomy, University of Southampton,  Southampton, SO17 1BJ, UK}

$^3${Department of Physics \& Astronomy, University College London, Gower Street, London, WC1E 6BT, UK}

$^4${The Oskar Klein Centre, Department of Physics, Stockholm University, SE - 106 91 Stockholm, Sweden}

$^5${Department of Astronomy and Astrophysics, University of Chicago, Chicago, IL 60637, USA}

$^6${Kavli Institute for Cosmological Physics, University of Chicago, Chicago, IL 60637, USA}

$^7${School of Mathematics and Physics, University of Queensland,  Brisbane, QLD 4072, Australia}

$^8${Department of Physics, Duke University Durham, NC 27708, USA}

$^9${Department of Physics, University of Oxford, Denys Wilkinson Building, Keble Road, Oxford OX1 3RH, United Kingdom}

$^{10}${Department of Physics, Duke University Durham, NC 27708, USA}

$^{11}${The Research School of Astronomy and Astrophysics, Australian National University, ACT 2601, Australia}

$^{12}${Institute of Astronomy and Kavli Institute for Cosmology, Madingley Road, Cambridge CB3 0HA, UK}

$^{13}${Center for Astrophysics $\vert$ Harvard \& Smithsonian, 60 Garden Street, Cambridge, MA 02138, USA}

$^{14}${Fermi National Accelerator Laboratory, P. O. Box 500, Batavia, IL 60510, USA}

$^{15}${Institute of Space Sciences (ICE, CSIC),  Campus UAB, Carrer de Can Magrans, s/n,  08193 Barcelona, Spain}

$^{16}${Institut d'Estudis Espacials de Catalunya (IEEC), 08034 Barcelona, Spain}

$^{17}${Department of Physics and Astronomy, Baylor University, One Bear Place 97316, Waco, TX 76798-7316, USA}

$^{18}${Universit\'e Grenoble Alpes, CNRS, LPSC-IN2P3, 38000 Grenoble, France}

$^{19}${Department of Physics and Astronomy, University of Pennsylvania, Philadelphia, PA 19104, USA}

$^{20}${Centre for Astrophysics \& Supercomputing, Swinburne University of Technology, Victoria 3122, Australia}

$^{21}${Department of Physics, Lancaster University, Lancs LA1 4YB, UK}

$^{22}$Institute of Space Sciences (ICE, CSIC),  Campus UAB, Carrer de Can Magrans, s/n,  08193 Barcelona, Spain

$^{23}$Physik-Institut, University of Zürich, Winterthurerstrasse 190, CH-8057 Zürich, Switzerland

$^{24}$Centro de Investigaciones Energ\'eticas, Medioambientales y Tecnol\'ogicas (CIEMAT), Madrid, Spain

$^{25}$Institute of Cosmology and Gravitation, University of Portsmouth, Portsmouth, PO1 3FX, UK

$^{26}$ Department of Physics, Northeastern University, Boston, MA 02115, USA

$^{27}$ University Observatory, LMU Faculty of Physics, Scheinerstr. 1, 81679 Munich, Germany

$^{28}$ Kavli Institute for Particle Astrophysics \& Cosmology, P. O. Box 2450, Stanford University, Stanford, CA 94305, USA  

$^{29}$ SLAC National Accelerator Laboratory, Menlo Park, CA 94025, USA

$^{30}$ Instituto de Astrofisica de Canarias, E-38205 La Laguna, Tenerife, Spain 

$^{31}$ Laborat\'orio Interinstitucional de e-Astronomia - LIneA, Av. Pastor Martin Luther King Jr, 126 Del Castilho, Nova Am\'erica Offices, Torre 3000/sala 817 CEP: 20765-000, Brazil

$^{32}$  Institut de F\'{\i}sica d'Altes Energies (IFAE), The Barcelona Institute of Science and Technology, Campus UAB, 08193 Bellaterra (Barcelona) Spain

$^{33}$ Oxford College of Emory University, Oxford, GA 30054, USA

$^{34}$  Hamburger Sternwarte, Universit\"{a}t Hamburg, Gojenbergsweg 112, 21029 Hamburg, Germany

$^{35}$  California Institute of Technology, 1200 East California Blvd, MC 249-17, Pasadena, CA 91125, USA

$^{36}$ Instituto de Fisica Teorica UAM/CSIC, Universidad Autonoma de Madrid, 28049 Madrid, Spain

$^{37}$ Santa Cruz Institute for Particle Physics, Santa Cruz, CA 95064, USA

$^{38}$ Center for Cosmology and Astro-Particle Physics, The Ohio State University, Columbus, OH 43210, USA 

$^{39}$, Department of Physics, The Ohio State University, Columbus, OH 43210, USA

$^{40}$ Department of Physics, University of Michigan, Ann Arbor, MI 48109, USA

$^{41}$ Center for Astrophysics $\vert$ Harvard \& Smithsonian, 60 Garden Street, Cambridge, MA 02138, USA

$^{41}$ Australian Astronomical Optics, Macquarie University, North Ryde, NSW 2113, Australia 

$^{42}$ Lowell Observatory, 1400 Mars Hill Rd, Flagstaff, AZ 86001, USA

$^{43}$ Jet Propulsion Laboratory, California Institute of Technology, 4800 Oak Grove Dr., Pasadena, CA 91109, USA

$^{43}$ Centre for Gravitational Astrophysics, College of Science, The Australian National University, ACT 2601, Australia

$^{44}$ George P. and Cynthia Woods Mitchell Institute for Fundamental Physics and Astronomy, and Department of Physics and Astronomy, Texas A\&M University, College Station, TX 77843,  USA

$^{45}$ Center for Astrophysical Surveys, National Center for Supercomputing Applications, 1205 West Clark St., Urbana, IL 61801, USA 

$^{46}$ Department of Astronomy, University of Illinois at Urbana-Champaign, 1002 W. Green Street, Urbana, IL 61801, USA

$^{47}$ Instituci\'o Catalana de Recerca i Estudis Avan\c{c}ats, E-08010 Barcelona, Spain 

$^{48}$ Institut de F\'{\i}sica d'Altes Energies (IFAE), The Barcelona Institute of Science and Technology, Campus UAB, 08193 Bellaterra (Barcelona) Spain

$^{49}$ Department of Physics, University of Cincinnati, Cincinnati, Ohio 45221, USA 

$^{50}$ Perimeter Institute for Theoretical Physics, 31 Caroline St. North, Waterloo, ON N2L 2Y5, Canada

$^{51}$ Department of Astrophysical Sciences, Princeton University, Peyton Hall, Princeton, NJ 08544, USA

$^{52}$ Centro de Tecnologia da Informa\c{c}\~ao Renato Archer, Campinas, SP, Brazil - 13069-901 

$^{53}$ Observat\'orio Nacional, Rio de Janeiro, RJ, Brazil - 20921-400

$^{54}$ Ruhr University Bochum, Faculty of Physics and Astronomy, Astronomical Institute, German Centre for Cosmological Lensing, 44780 Bochum, Germany

$^{55}$  Nordita, KTH Royal Institute of Technology and Stockholm University, Hannes Alfv\'ens v\"ag 12, SE-10691 Stockholm, Sweden

$^{56}$ School of Mathematics and Physics, University of Surrey, Guildford, Surrey, UK

$^{57}$  Department of Physics and Astronomy, Pevensey Building, University of Sussex, Brighton, BN1 9QH, UK

$^{58}$  Computer Science and Mathematics Division, Oak Ridge National Laboratory, Oak Ridge, TN 37831

$^{59}$  Cerro Tololo Inter-American Observatory, NSF's National Optical-Infrared Astronomy Research Laboratory, Casilla 603, La Serena, Chile

$^{60}$ Berkeley Center for Cosmological Physics, Department of Physics, University of California, Berkeley, CA 94720

$^{61}$ US Lawrence Berkeley National Laboratory, 1 Cyclotron Road, Berkeley, CA 94720, USA

$^{62}$ INAF-Osservatorio Astronomico di Trieste, via G. B. Tiepolo 11, I-34143 Trieste, Italy, Laboratorio Interinstitucional de e-Astronomia - LIneA, Av. Pastor Martin Luther King Jr, 126 Del Castilho, Nova America Offices, Torre 3000/sala 817, Brazil



\bibliographystyle{mnras}
\bibliography{research2, matt} 


\bsp	
\label{lastpage}
\end{document}

%% file: SYSUNCER_VALS.tex
\def\BSTW{0.029}
\def\BSTWDEL{-0.007}
\def\BSTWPERC{5.6}
\def\PSYSONE{0.006}
\def\PSYSONEDEL{0.001}
\def\PSYSONEPERC{1.1}
\def\PSYSTWO{0.021}
\def\PSYSTWODEL{-0.004}
\def\PSYSTWOPERC{4.1}
\def\PSYSTHREE{0.025}
\def\PSYSTHREEDEL{-0.009}
\def\PSYSTHREEPERC{4.8}
\def\WISEMAN{0.02}
\def\WISEMANDEL{-0.008}
\def\WISEMANPERC{3.8}
\def\MWEBV{0.024}
\def\MWEBVDEL{-0.011}
\def\MWEBVPERC{4.5}
\def\CALIB{0.066}
\def\CALIBDEL{-0.073}
\def\CALIBPERC{12.6}
\def\CALSPEC{0.009}
\def\CALSPECDEL{-0.001}
\def\CALSPECPERC{1.8}
\def\ZSHIFT{0.011}
\def\ZSHIFTDEL{-0.0}
\def\ZSHIFTPERC{2.2}
\def\MASSLOC{0.0}
\def\MASSLOCDEL{0.0}
\def\MASSLOCPERC{0.0}
\def\ALPHAEVOL{0.013}
\def\ALPHAEVOLDEL{-0.002}
\def\ALPHAEVOLPERC{2.6}
\def\BETAEVOL{0.0}
\def\BETAEVOLDEL{-0.001}
\def\BETAEVOLPERC{0.0}
\def\GAMMAEVOL{0.003}
\def\GAMMAEVOLDEL{-0.0}
\def\GAMMAEVOLPERC{0.6}
\def\SNN{0.01}
\def\SNNDEL{-0.001}
\def\SNNPERC{2.0}
\def\SNIRF{0.002}
\def\SNIRFDEL{0.0}
\def\SNIRFPERC{0.3}
\def\SCONE{0.003}
\def\SCONEDEL{-0.0}
\def\SCONEPERC{0.6}
\def\CCPRIOR{0.003}

\def\CCPRIORPERC{0.6}
\def\COLORLAW{0.017}
\def\COLORLAWDEL{-0.006}
\def\COLORLAWPERC{3.3}
\def\FIXAB{0.0}
\def\FIXABDEL{0.0}
\def\FIXABPERC{0.0}
\def\INTRSCCOLOR{0.0}
\def\INTRSCCOLORDEL{0.029}
\def\INTRSCCOLORPERC{0.0}
\def\SIGINT{0.007}
\def\SIGINTDEL{-0.0}
\def\SIGINTPERC{1.3}
\def\SVAHOST{0.0}
\def\SVAHOSTDEL{0.006}
\def\SVAHOSTPERC{0.0}
\def\HOSTEFF{0.002}
\def\HOSTEFFDEL{0.0}
\def\HOSTEFFPERC{0.4}
\def\VPEC{0.029}
\def\VPECDEL{-0.016}
\def\VPECPERC{5.6}

%% file: SystUncertainty.tex
\begin{table}
\centering
\begin{tabular}{lclcc}
\hline
Systematic & $\sigma_{w}$ & \%(tot) & $\delta w$ & \citetalias{DES5YR} $\sigma_{w}$\\
\hline
\textbf{Total Stat+Syst} & $0.142$ & 100 & 0 & $0.152$ \\
\textbf{Total Statistical} & $0.11$ & N/A & $-0.092$ & $0.132$ \\
\hline \vspace{-2mm}\\
\textbf{Calibration $\&$ LC model} & \textbf{0.075} & 14.4\% & -- & \textbf{0.057} \\
\hspace{2mm} SALT3+Calibration & \CALIB & \CALIBPERC & \CALIBDEL & $0.052$ \\
\hspace{2mm} HST Calspec & \CALSPEC & \CALSPECPERC & \CALSPECDEL & $0.006$ \\
\\
\textbf{SN Ia astrophysics} & \textbf{0.124} & 23.9\% & -- & \textbf{0.133} \\
\hspace{2mm} P23 dust pop 1 &  \PSYSONE & \PSYSONEPERC & \PSYSONEDEL & $0.019$ \\
\hspace{2mm} P23 dust pop 2 &  \PSYSTWO & \PSYSTWOPERC & \PSYSTWODEL & $0.024$ \\
\hspace{2mm} P23 dust pop 3 &  \PSYSTHREE & \PSYSTHREEPERC & \PSYSTHREEDEL & $0.020$ \\
\hspace{2mm} P23($u-r$) &  \INTRSCCOLOR & \INTRSCCOLORPERC & \INTRSCCOLORDEL & $0.00$ \\
\hspace{2mm} Dust model as in \citetalias{BS20}  &  \BSTW & \BSTWPERC & \BSTWDEL & $0.027$ \\
\hspace{2mm} Model SN age (W22) &  \WISEMAN & \WISEMANPERC & \WISEMANDEL & $0.00$ \\
\hspace{2mm} Fixed $\alpha \beta$  &  \FIXAB & \FIXABPERC & \FIXABDEL & $0.002$ \\
\hspace{2mm} $\alpha$ Evolution &  \ALPHAEVOL & \ALPHAEVOLPERC & \ALPHAEVOLDEL & $0.020$ \\
\hspace{2mm} $\beta$ Evolution & \BETAEVOL & \BETAEVOLPERC & \BETAEVOLDEL  & $0.000$ \\
\hspace{2mm} $\gamma$ Evolution &  \GAMMAEVOL & \GAMMAEVOLPERC & \GAMMAEVOLDEL &  $0.011$ \\
\hspace{2mm} Mass step location &  \MASSLOC & \MASSLOCPERC & \MASSLOCDEL & $0.000$ \\
\hspace{2mm} $\sigma_{\mathrm{int}}$ modeling &  \SIGINT & \SIGINTPERC & \SIGINTDEL & $0.013$ \vspace{2mm}\\
\textbf{Milky Way extinction} & \textbf{0.041} & 7.7\% & -- & \textbf{0.034} \\
\hspace{2mm} MW 5$\%$ scaling  & \MWEBV & \MWEBVPERC & \MWEBVDEL  & $0.020$ \\
\hspace{2mm} MW colour law CCM  & \COLORLAW & \COLORLAWPERC & \COLORLAWDEL  & $0.014$ \vspace{2mm}\\
\textbf{Survey modeling } & \textbf{0.002} & 0.4\% & -- & \textbf{0.015} \\
\hspace{2mm} DES SV catalog & \SVAHOST & \SVAHOSTPERC & \SVAHOSTDEL & $0.009$ \\
\hspace{2mm} Shift $\epsilon_{z}^{\mathrm{spec}}$ & \HOSTEFF & \HOSTEFFPERC & \HOSTEFFDEL & $0.005$ \vspace{2mm}\\
\textbf{Contamination} & \textbf{0.018} & 3.5\%  & -- & \textbf{0.028} \\
\hspace{2mm} Classifier SCONE  & \SCONE & \SCONEPERC & \SCONEDEL & $0.006$ \\
\hspace{2mm} Classifier SNIRF & \SNIRF & \SNIRFPERC & \SNIRFDEL & $0.013$ \\
\hspace{2mm} SuperNNova different training  & \SNN & \SNNPERC & \SNNDEL & $0.006$ \\
\hspace{2mm} Core-collapse SN prior   & \CCPRIOR & \CCPRIORPERC & \SNNDEL & $0.003$ \vspace{2mm}\\
\textbf{Redshift} & \textbf{0.040} & 7.8\% & -- & \textbf{0.037} \\
\hspace{2mm} Redshift shift & \ZSHIFT & \ZSHIFTPERC & \ZSHIFTDEL & $0.012$  \\
\hspace{2mm} Peculiar velocities & \VPEC & \VPECPERC & \VPECDEL & $0.025$ \\
\end{tabular}
\caption{Systematic uncertainties considering SN-only, without a CMB prior. More detail on each systematic is given in Section \ref{sec:SYST} and Table \ref{tab:syst_description}. Bolded numbers are the sum of each individual component. For comparison, we present the $\sigma_w$ values from \citetalias{DES5YR}. Unlike in Table \ref{tab:systematic_uncertainties}, the $\delta w$ presented here use the full covariance matrix as presented in Section \ref{sec:Methodology}.}
\label{tab:syst_size}
\end{table}
